\documentclass[aps,amsmath,amssymb,nofootinbib,preprintnumbers]{revtex4} 
\usepackage{graphicx} 
\usepackage{amsmath}
\usepackage{amsfonts,amsbsy}
\usepackage{amssymb}
\usepackage{color}
\usepackage{mathtools}

\newcommand{\dimer}[4]{\ensuremath{   \langle \rho^{#1} (#2) \rho^{#3} (#4) \rangle_P }}

\newcommand{\phimerM}[4]{\ensuremath{  \phi^{#1} (#2, #3; #4) }}

\def\be{\begin{equation}}
\def\ee{\end{equation}}
\def\bea{\begin{eqnarray}}
\def\eea{\end{eqnarray}}

\def\beq{\begin{equation}}
\def\eeq{\end{equation}}

\newcommand{\nn}{\nonumber}

\begin{document}


\title{ 
Initial state $qqg$ correlations as a background for the Chiral Magnetic Effect  in collision of small systems.
}

\author{Alex Kovner}
\affiliation{Physics Department, University of Connecticut, 2152 Hillside Road, Storrs, CT 06269, USA
\\
 Departamento de F\'\i sica, Universidad T\'ecnica
Federico Santa Mar\'\i a, Avda. Espa\~na 1680,
Casilla 110-V, Valparaiso, Chile
}

\author{Michael Lublinsky}
\affiliation{Physics Department, Ben-Gurion University of the Negev, Beer Sheva 84105, Israel}

\author{Vladimir Skokov}
\affiliation{RIKEN/BNL, Brookhaven National Laboratory, Upton, NY 11973}

\begin{abstract}
Motivated by understanding  the background to  Chiral Magnetic Effect in proton-nucleus collisions from first principles, 
 we compute the three particle correlation in the projectile wave function.
We extract the correlations between two quarks and one gluon
in the framework of the Color Glass Condensate. This is related to the 
same-charge correlation of the conventional observable for the  Chiral Magnetic Effect.  
We show that there are two different contributions to this correlation function. One contribution is rapidity-independent  and as such can be identified with the pedestal; 
while the other displays rather strong rapidity dependence.
The pedestal contribution and the rapidity-dependent contribution at large rapidity separation between the two quarks  result
in the negative same charge correlations, while at small rapidity separation the second contribution changes sign. We argue that the computed initial state correlations might 
be partially responsible for the experimentally observed signal in proton-nucleus collisions.  
\end{abstract}

\date{\today} 
\preprint{RBRC 1242}

\maketitle

\section{Introduction} 

Topological fluctuations in the early time Glasma state~\cite{Mace:2016svc,Mace:2016shq} or 
thermal sphaleron transitions~\cite{Arnold:1987mh,Moore:2010jd} during later stages of heavy-ion collisions may lead to  
the Chiral Magnetic Effect (CME)~\cite{Kharzeev:2007jp}, the generation of the electromagnetic 
current along the magnetic field~\cite{Skokov:2009qp,Bzdak:2011yy,McLerran:2013hla}. 
Experimentally, the associated charge separation can be  measured by three particle angular average~\cite{Voloshin:2004vk}, 
$\gamma = \langle \cos( \phi_\alpha + \phi_\beta - 2\phi ) \rangle$, where $\phi$ is the azimuthal angle of a trigger
particle defining  the reaction plane and $\phi_{\alpha,\beta}$ are azimuthal angles of associate particles carrying electric charge. 
The averaging is usually taken over a range of transverse momenta of the charged particles. The observable 
$\gamma$ is also often considered as a function of relative rapidity separation between the charge particles $\Delta \eta$  and
the multiplicity/centrality of the collision. The charge of the particles $\alpha$ and $\beta$ can be either same or opposite. 
For more information about the experimental measurements, see Ref.~\cite{Voloshin:2004vk,Abelev:2009ac,Abelev:2009ad,Khachatryan:2016got,Skokov:2016yrj,Tribedy:2017hwn}.  

The CME prediction for the observable $\gamma$ can be understood as follows. In the presence of the
strong magnetic field, $B$,  and initial axial charge $\mu_5$, the CME builds an electric current along the 
magnetic field~\cite{Kharzeev:2007jp}, which in non-central heavy-ion collisions points in the 
out-of-plane directions, see Refs.~\cite{Skokov:2009qp,Bzdak:2011yy}. The current results in the 
transport of the charges and subsequent formation of a dipole moment in the charge distribution, which can be described by 
\begin{equation}
	\frac{ dN_\alpha} 
	{d \phi}  = {\cal N} \left( 
	1 + 
	2 v_1 \cos(\phi - \psi_{\rm RP})
	+ 
	2 v_2 \cos(2[\phi - \psi_{\rm RP}])
	+ 2 a_\alpha \sin(\phi - \psi_{\rm RP}) + \dots 
	\right) ,
\end{equation}
where $\psi_{\rm RP}$ is the reaction plane angle (neglecting the fluctuations, the magnetic field is perpendicular to the reaction plane), 
$v_1$ is the directed flow, $v_2$ is the elliptic flow and $\alpha=+,-$ denotes the charge of the particles. The parameters 
$a_\pm$ describe the formation of the electric dipole $a_+ = - a_- \propto \mu_5 B$. The sign of $\mu_5$  fluctuates 
on event by event basis rendering $\langle a_\pm \rangle =0$. Nevertheless, the parity-even  fluctuations,  $\langle a_\alpha a_{\alpha'} \rangle$,   
can still be measured in experiment. The observable $\gamma$ suppresses the  background~\cite{Voloshin:2004vk}
and is approximately equals to the fluctuations, that is   
\begin{equation}
	\gamma \approx \langle -  a_{\alpha} a_{\alpha'} \rangle .  
\end{equation}
From this expression one can draw conclusions on the CME predictions for $\gamma$. For the same charges $\alpha '= \alpha $,  $a_{\alpha'} =  a_{\alpha}$ and
thus one expects  $\gamma \approx - \langle a^2 \rangle  < 0$; for opposite charges  $\alpha' = - \alpha$, $a_{\alpha'} =  -  a_{\alpha}$  
and thus $\gamma \approx \langle a^2 \rangle  > 0$. Additionally, if the backgrounds effects are negligible,  the 
same-charge correlator $\gamma$ should be opposite in sign but equal in magnitude to opposite-sign correlator.

In collision with heavy-ions, the first measurement of $\gamma$ were performed at RHIC~\cite{Abelev:2009ac,Abelev:2009ad};
it was observed that opposite-charge correlations were very close to zero or even negative, while 
the same-charge were negative and larger in the amplitude. The observation of close to zero opposite-charge correlations was not immediately  
consistent with CME, as it was predicted to have the same amplitude as same-charge. However,
the observable $\gamma$ might be potentially contaminated by large charge-independent backgrounds,  that  
shift the values of both the same-charge and opposite-charge correlations. 

In order to test CME, a few other measurements and observables were explored, for details see Ref.~\cite{Kharzeev:2015znc,Skokov:2016yrj}. 
Nevertheless, the status of  the CME in heavy-ion collisions  remains inconclusive
due to background correlations that may be responsible for the entirety of the observed signal~\cite{Schlichting:2010qia,Pratt:2010zn,Bzdak:2010fd,Bzdak:2012ia}.  

Recently, the CMS collaboration performed measurements of the three particle correlations  in proton-nucleus collisions~\cite{Khachatryan:2016got} 
at $\sqrt s $ = 5 TeV.    
The CME predicts virtually absent signal in p-A collisions due to small values of the magnetic field and its decorrelation
with the event plane. However, it was  observed that the differences between the same and opposite sign correlations,
as functions of multiplicity and rapidity gap between the two charged particles,
are of a similar magnitude in proton-nucleus and nucleus-nucleus collisions at the same multiplicities. 
This does not necessarily pose an immediate challenge to the CME interpretation of the charge dependent 
azimuthal correlations in heavy ion collisions, as the results coincide 
only in peripheral bins of  Pb-Pb collisions, where the background effects are  expected to play a  dominant role~\cite{Tribedy:2017hwn}.  
 
Nevertheless, the CMS measurements make it clear that without microscopic understanding of the background contributions to 
the observable $\gamma$, any interpretation of the data will be unsatisfactory.  
Motivated by the data of the CMS collaboration, in this paper, we address one
of the possible sources of this background; we concentrate on the same-charge correlations, which usually fall outside the scope 
of the conventional background models~\cite{Schlichting:2010qia,Hirono:2014oda}
except for the global transverse momentum conservation. We work in the
framework of the Color Glass Condensate; which 
was successful in predicting  the ``ridge'' correlations and is often utilized to address the systematics of the 
azimuthal anisotropy in the initial state, see
Refs.~\cite{Dumitru:2010iy,Kovner:2010xk,Kovner:2012jm,Kovchegov:2012nd,Dusling:2012iga,Dusling:2013qoz,Dumitru:2014yza,Skokov:2014tka,McLerran:2016snu,Lappi:2015vha,Schenke:2015aqa,Dumitru:2015cfa,Altinoluk:2015uaa,Lappi:2015vta,McLerran:2015sva,Schlichting:2016sqo,Dusling:2017dqg,Gotsman:2017zoq}.   

As was shown in Ref.~\cite{Altinoluk:2015uaa} the Bose-Einstein enhancement (BSE) of soft gluons in the projectile provides the physical interpretation of the glasma-graph calculation of the 
``ridge'' correlations. In a follow up paper the authors of Ref.~\cite{Altinoluk:2016vax} also explored the consequences of quark's Pauli blocking 
in the projectile wave function.  Mindful of these studies, we consider the observable $\gamma$ and explore possible nontrivial contribution to this observable, and therefore the CME background  stemming from the quantum correlations in the initial state.   In practice, 
we consider the angular average  $\gamma$ defined as a projectile average $\gamma= \langle \cos (\phi_p + \phi_q -2 \phi_m) \rangle $
where $p$ and $q$ are the transverse  momenta of two same-charge/same-flavor quarks in the  light cone wave-function of the projectile, 
and $m$ is the transverse momentum of the gluon.  We will demonstrate that there are two distinct contributions to this quantity: the pedestal, the rapidity-independent
contribution, with a negative $\gamma$, and the rapidity-dependent and sign changing contribution originating from Pauli blocking.

The paper is organized as follows. In Sec.~\ref{Sec:Pre} we briefly review the relevant results of Ref.~\cite{Altinoluk:2016vax} for quark-quark correlations, originating from two quark-antiquark pairs in the wave function. In Sec.~\ref{sec:quark} we extend this calculation to include three particle correlations, 
computing contribution of an additional gluon  thus bringing up the relevant Fock state component to 5 particles. 
In this paper we limit ourselves to the calculation of correlations in the wave function of the incoming hadron, and do not attempt to calculate three particle production, which we leave for future work. Nevertheless, as demonstrated in \cite{Altinoluk:2015uaa, Altinoluk:2016vax} such initial state correlations within the CGC approach have a direct effect on production of particles, and thus can serve as a basis for qualitative understanding of the effect.
In Sec.~\ref{Sec:Dis}
we  discuss and summarize our findings.  

\section{Preliminaries: Quark contribution to the projectile wave-function} 
\label{Sec:Pre}

Let  $d^\dagger$ and $d$ denote quark creation and annihilation operators,
while $\bar d^\dagger$ and $\bar d$ are those of the antiquark. 
Additionally for gluons, we introduce $a^+$ and $a$. 

First we formulate,  two particle, quark-antiquark, content of the light-cone wave function.
This will allow us to introduce the notation we use when considering a more complicated case 
of two quark-two antiquark and gluon. 
In perturbative calculations, the quarks and antiquarks appear in the light-cone
wave function of a valence charge  via soft-gluon splitting or instantaneous interaction,
see details in Ref.~\cite{Altinoluk:2016vax}, Appendix A and the review~\cite{Kogut:1969xa,Bjorken:1970ah,Brodsky:1997de}. 
The quark-antiquark component of the
light cone wave function of a ``dressed'' color charge density is given
by\footnote{The state to this order in perturbation theory
contains  one-gluon and two-gluon components. We do not indicate those
explicitly, as they do not contribute to correlated quark-gluon production. 
These contributions were studied, e.g. in Refs.~\cite{Altinoluk:2015uaa}. 
}
\bea
\label{vd}
|v\rangle^D_2
&=&
(1\,-\,g^4\,\kappa_4)\,|v\rangle
\nonumber \\
&+&
	g^2\,
\int {dk^+d\alpha\,d^2p\,d^2q\over (2\pi)^3}\
\zeta^{\gamma\delta}_{s_1 s_2}(k^+,p,q,\alpha)\ d^{\dagger \gamma}_{s_1}(q^+,q)\,
\bar d_{s_2}^{\dagger\delta}(p^+,p)
\,
\,|v\rangle,
\eea
where $|v\rangle$ denotes a valence state characterised by a distribution of  charge densities $\rho$ of valence (fast) partons. 
The subscript ``2'' in  $|v\rangle_2$   counts the perturbative order in  the Yang-Mills coupling denoted as $g$.
$\kappa_4$ is a  constant ensuring the correct
normalisation of the dressed state,  $\gamma,\delta = 1, 2,\dots, N_c$ are fundamental color indices, and $s_{1,2}$ stand for the spinor indices.
The value of $\kappa_4$ is irrelevant for the problem at hand.
We define the longitudinal momentum fraction $\alpha$  as
\beq
 p^+ = \alpha k^+,\ \
q^+ = \bar \alpha k^+, \ \  \bar\alpha =1-\alpha,
\eeq
with $k$ the momentum of the parent gluon that splits into a quark and an antiquark. The splitting amplitude $\zeta$ is given by
\begin{equation}
\label{zeta_def}
\zeta^{\gamma\delta}_{s_1 s_2}(k^+,p,q,\alpha)\ =
\tau^a_{\gamma\delta}\, \int \frac{d^2k}{(2\pi)^2}\,\rho^a(k)\ \phi_{s_1 s_2}(k,p,q;\alpha),
\end{equation}
where $\tau^a$ are the generators of $SU(N_c)$ in the fundamental representation. Here, 
\beq
\phi\,=\,\phi^{(1)}\ +\ \phi^{(2)},
\eeq
with
\begin{equation}
 \phi^{(1)}_{s_1 s_2}(k,p,q;\alpha) =-\delta_{s_1s_2}\frac{2\alpha\bar\alpha}{\bar\alpha p^2+\alpha q^2}(2\pi)^2 \delta^{(2)}(k-p-q)
\end{equation}
and
\bea
\phi^{(2)}_{s_1s_2}(k,p,q;\alpha)
=
\frac{1 }{ k^2\,\left[\bar\alpha p^2\,+\,
\alpha q^2\right]}
\left\{2\alpha\bar\alpha k^2-\left(\bar\alpha k\cdot p+\alpha k\cdot q\right)+2i\sigma^3k\times p\right\} (2\pi)^2\delta^{(2)}(k-p-q).
\eea
Thus,
\begin{equation}
\phi_{s_1s_2}(k,p,q;\alpha)=\phi_{s_1s_2}(k,p;\alpha) (2\pi)^2\delta^{(2)}(k-p-q)
\end{equation}
where 
\bea
\phi_{s_1s_2}(k,p;\alpha)=
\frac{1 }{ k^2\,\left[\bar\alpha p^2\,+\,
\alpha (k-p)^2\right]} 
\Big\{-\left[\bar\alpha k\cdot p+\alpha k\cdot (k-p)\right]+2i\sigma^3k\times p\Big\}.
\eea
The $\phi^{(1)}$ term comes from the instantaneous interaction, while
$\phi^{(2)}$ from the soft gluon splitting.

However, $|v\rangle^D_2$ is not the state we are interested in, as it provides 
information about quark-antiquark content of the light cone wave function only.  
To probe quark-quark-gluon correlations we have to consider the two quark--two
antiquark and gluon component of the dressed  state, that is the state with 5 particles. 
We will adopt the same strategy as was used in the glasma graph calculation.
That is, we focus on terms enhanced
by the charge density in the wave-function;  similar approach was also used 
in Ref.~\cite{Kovchegov:2012nd}.  
At the lowest order the relevant component of the wave function is given by
\begin{eqnarray}\label{v4}
	|v\rangle^D_5&=&{\rm virtual} \notag \\ &+& \frac{g^4}{2}\,
\int {dk^+d\alpha\,d^2p'\,d^2\bar{p}'\over (2\,\pi)^3}
{d\bar k^+d\beta\, d^2q'\, d^2\bar{q}'\over (2\,\pi)^3}  
\zeta^{\epsilon\iota}_{s'_1 s'_2}(k^+,p',\bar{p}';\alpha)
\zeta^{\gamma\delta}_{r_1 r_2}(\bar k^+,q',\bar{q}';\beta) \notag  \\ \notag  
&\times&\underbracket{ 
d_{s'_1}^{\dagger \epsilon}(\bar{\alpha}k^+, p')\,
\bar d_{s'_2}^{\dagger\iota}(\alpha k^+,\bar{p}')\ 
d_{r_1}^{\dagger\gamma}(\bar{\beta}\bar{k}^+, q')\,
\bar d_{r_2}^{\dagger\delta}(\beta\bar{k}^+,\bar{q}')
\hspace{4cm} }_{\rm q\bar{q}  q\bar{q}  } \\ &\times&
\underbracket{ g\int \frac{dm^+}{(m^+)^{1/2}} \frac{d^2 m}{(2\pi)^3} 
\frac{m_i}{m^2} \rho^a(-m)
a^{\dagger a}_i(m^+,m)}_{\rm g}
|v\rangle \,,
\end{eqnarray}
where we explicitly showed the part corresponding to the pair of quark-antiquark and the soft gluon. 
In the following section, we will use this dressed state to find 
the average number of two quark and a gluon “triplets” in the wave function. 



\section{Two-quark-gluon correlations}
\label{sec:quark}

In this section we compute the correlations between the quarks and the gluon in
the CGC wave function of the projectile.
To be able to make definitive statements about correlations between produced particles this calculation has to be supplemented by the analysis of particle production, as in principle  momentum distribution of produced particles is affected by momentum transfer from the target. Also scattering is not equally efficient  in putting on shell all partons in the incoming wave function. In particular partons with large transverse momentum are  emitted into the final state with smaller probability. Thus correlations between emitted particles are not identical to correlations between the partons in the projectile wave function.  However as was observed in Ref.~\cite{Altinoluk:2015uaa,Altinoluk:2016vax,Kovner:2016jfp}
this change mostly affects the {\it quantitative} features preserving the {\rm qualitative} pattern of the correlation. 
In this exploratory study we only compute the correlation in the projectile wave function and consider this to be a proxy to correlation between produced particles, at least if the transverse momenta of these particles are not too large. 
Already on this level, as we will demonstrate below, the calculations 
are non-trivial and require numerical integration. 

The aim of this section is to compute the average number of  two quark and a gluon ``triplets'' in the wave
function that is formally defined, see e.g. Ref.~\cite{Greiner:1998jw}, as
\bea
\label{Npair}
\notag &&{dN\over dp^+d^2pdq^+d^2q dm^+ d^2m}\,=
\\&&
\frac{1}{(2\pi)^6}\,\left\langle ^{D}_{5}\langle v|
d^{\dagger}_{\alpha,s_1}(p^+,p)d^{\dagger}_{\beta,s_2}(q^+,q)\,d_{\beta,s_2}(q^+,q) 
\,\,d_{\alpha,s_1}(p^+,p)\,
a^{\dagger f}_i(m^+,m)
a^{f}_i(m^+,m)
|v\rangle^D_5\, \right\rangle_P \; ,
\eea
i.e. first, we need to calculate the expectation value of the ``number of quark-gluon triplets''
in our dressed state $|v\rangle_5^D$, and then, average over the color
charge densities in the projectile. For the latter we use the McLerran-Venugopalan  (Gaussian) model~\cite{McLerran:1993ni,McLerran:1993ka}. 
This choice is somewhat restrictive and might potentially affect the result in a non universal way, especially at lower collision energies, where the 
odderon component becomes stronger.
As it will be clear below, the observable we are computing has six powers of the charge density $\rho$ and thus might be sensitive to the odderon.   
 In principle the model can be extended along the lines of Ref.~\cite{Jeon:2005cf} where the odderon is included in the averaging weight on the classical level.

\begin{figure}
	\centerline{\includegraphics[width=0.3\linewidth]{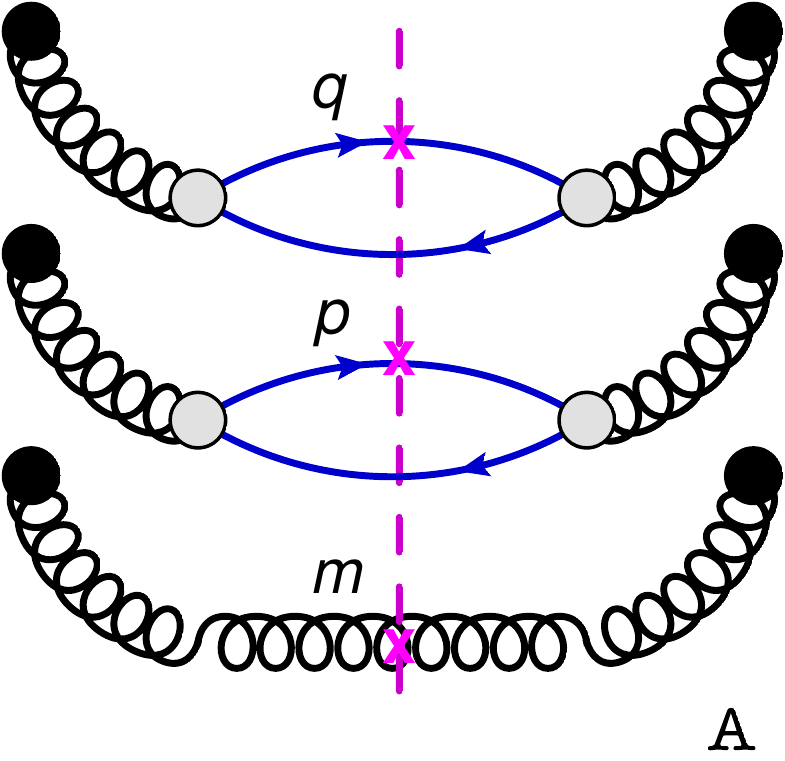}\hspace{2cm}
\includegraphics[width=0.3\linewidth]{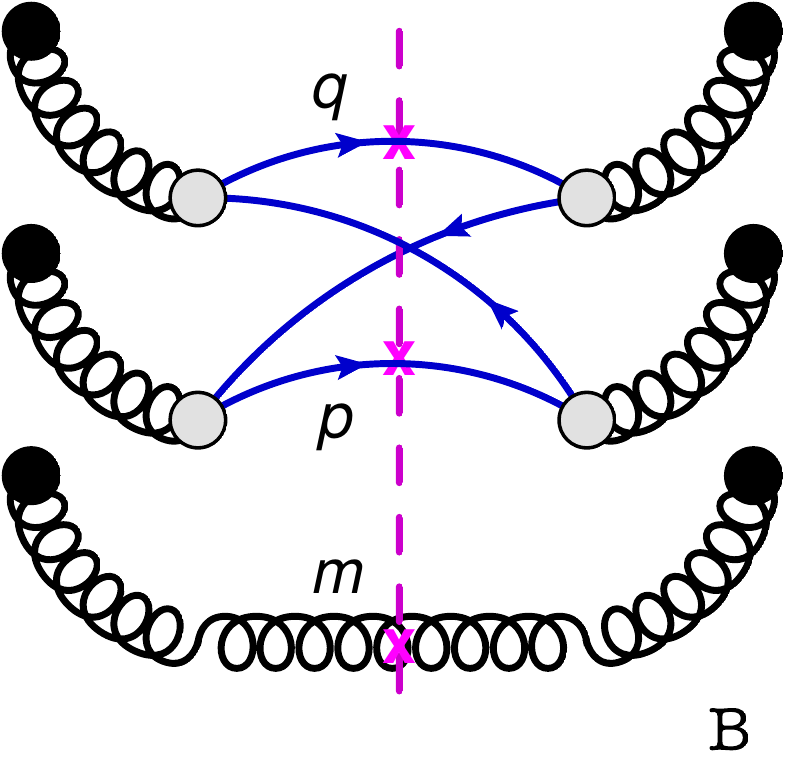}}
\caption{The diagrammatic representation of two distinct contributions in Eq.~\eqref{Npair_square_final}. 
The black blobs denote the 
gluon sources, $\rho$. The grey blob in the gluon splitting vertex accounts for the instantaneous interaction too. 
}
\label{fig:diagram}
\end{figure}

A similar problem was addressed  in Ref.~\cite{Altinoluk:2016vax} for two-quark correlations, 
see Appendix B of Ref.~\cite{Altinoluk:2016vax}. We need to extend this to include a gluon. This is, thankfully quite straightforward. The extra gluon is created in the wave function independently of the quark pair from the valence charge density. 
The resulting expression for the correlator is, see also Fig.~\ref{fig:diagram}, 
\bea
\label{Npair_square_final}
\frac{dN}{d\eta_1d^2pd\eta_2d^2q d\eta_g d^2m }&=&\frac{1}{(2\pi)^4}\frac{g^{10}}{m^2}\int 
d^2k \, d^2\bar{k} \, d^2l \, d^2\bar{l} \;
\langle \underbracket{\rho^a(k) \rho^c(\bar{k}) \rho^b(l) \rho^d(\bar{l})}_{\rm q \bar{q} q \bar{q}}
\underbracket{\rho^f(m) \rho^f(-m)}_{\rm g}
\rangle_P \nonumber\\
&&
\times
\Bigg\{ 
	\underbracket{{\rm tr}(\tau^a\tau^b) {\rm tr}(\tau^c\tau^d) \Phi_2(k,l; p) \Phi_2(\bar{k},\bar{l}; q)
}_{\mathtt A}
\underbracket {- {\rm tr}(\tau^a\tau^b\tau^c\tau^d) \Phi_4(k,l,\bar{k},\bar{l};p,q)
}_{\mathtt B}
\Bigg\}
,
\eea
where $\rho^a(k)$, $ \rho^b(\bar{k})$ and one of $\rho^f(m)$ are the color charge densities in the
amplitude and $\rho^c(l)$, $\rho^d(\bar{l})$ and the other $\rho^f(-m)$ are the color charge densities
in the complex conjugate amplitude. The rapidities are defined as
$\eta_1=\ln(m^+/p^+)$ and $\eta_2=\ln(m^+/q^+)$, with $m^+$ the gluon 
$+$-momentum. Note that the average number, Eq.~\eqref{Npair_square_final}, is independent of the gluon rapidity, $\eta_g$. 
The functions $\Phi_2$ and $\Phi_4$ are defined
respectively as \cite{Altinoluk:2016vax}
\beq
\label{Phi_2_def}
\Phi_2(k,l; p)\,\equiv\,\int_0^1 \,d\alpha\int \frac{d^2\bar{p}'}{(2\pi)^2}\sum_{s_1s_2}\
\phi_{s_1,s_2}(k,p,\bar{p}';\alpha)\ \phi_{s_1,s_2}^*(l,p,\bar{p}';\alpha)
\eeq
and
\bea
\label{Phi_4_def}
\Phi_4(k,l,\bar k,\bar l;p,q)&\equiv& \sum_{s_1,s_2,\bar s_1,\bar s_2} \int_0^1 \,
{d\alpha\,d\beta\over (\beta+\bar\beta e^{\eta_1-\eta_2})  (\alpha+\bar\alpha e^{\eta_2-\eta_1})}\,  \\
&&\hspace{-1.5cm}
\times\int \frac{d^2\bar{p}'}{(2\pi)^2}\frac{d^2\bar{q}'}{(2\pi)^2} 
\phi_{s_1s_2}(k,p,\bar{p}';\alpha)\ 
\phi_{\bar s_1\bar s_2}(\bar k,q,\bar{q}';\beta)\
\phi_{s_1\bar s_2}^*(l,p,\bar{q}';\beta)\,
\phi_{\bar s_1s_2}^*(\bar l,q,\bar{p}';\alpha). \nn
\eea
The integrals with respect to prime momenta represent ``inclusiveness'' over the antiquarks. 
The integrals over $\bar {p}', \ \bar {q}'$ reduce the number of $\delta$-functions to two, so that in general we can write
\bea
\Phi_4(k,l,\bar k,\bar l;p,q)&=& \sum_{s_1\,s_2,\bar s_1,\bar s_2} \int_0^1 \,
{d\alpha\,d\beta\over (\beta+\bar\beta e^{\eta_1-\eta_2})  (\alpha+\bar\alpha e^{\eta_2-\eta_1})}
\,
 \\
&\times&
\phi_{s_1s_2}(k,p;\alpha)\; \phi_{\bar s_1\bar s_2}(\bar k,q;\beta)\;  \phi_{s_1\bar s_2}^*(\bar k-q+p,p;\beta)\;
\phi_{\bar s_1s_2}^*(k+q-p,q;\alpha) \nn \\
&\times&
(2\pi)^2\delta^{(2)}(\bar l-k-q+p)\; (2\pi)^2\delta^{(2)}(l-\bar k+q-p).\nn
\eea

Lets  comment on the origin of different terms in Eq.~\eqref{Npair_square_final}. First, the particle density is proportional to 
$g^{10}$: two powers of the coupling constant come from the gluon production, and the leftover $g^8$ originate from production  of two quark-antiquark pairs, as each is proportional to $g^4$ owing to production of a gluon and its splitting into a quark-antiquark pair. The gluon component of    
Eq.~\eqref{Npair_square_final} is trivial and proportional to $g^2 \rho(m) \rho(-m)/m^2$, or the square of the Weizs\"acker-Williams 
field $ig \rho_a(m) m^j/m^2$ . 
The quark contribution coincides with that 
of Ref.~\cite{Altinoluk:2016vax}. It contains two distinct  contributions in the curly brackets of  Eq.~\eqref{Npair_square_final}:
($\mathtt{A}$) the term proportional to $\Phi_2$ corresponds to two quark loops and thus contributes with the positive sign, while the term 
($\mathtt{B}$) 
proportional to $\Phi_4$ has one quark loop resulting in the minus sign, see Fig.~\ref{fig:diagram}. 
The latter term manifests the Pauli blocking; with the minus 
sign leading to the dilution of the correlation!

\subsection*{The Pauli blocking term} 

As it is clear from Eq.~\eqref{Npair_square_final} 
the correlations between the gluon and quarks originates only from the 
averaging over the projectile color densities. In a Gaussian model for 
the projectile, thus we will not consider the  
terms involving the contraction $\langle \rho^f(m) \rho^f(-m) \rangle_P$, since this contraction leads to uncorrelated gluon production.
Additionally, we will postpone the consideration of the first term in 
the curly brackets of Eq.~\eqref{Npair_square_final}. This term contributes to the correlated quark production only in subleading order at  large $N_c$. 
Nevertheless compared to the second term of Eq.~\eqref{Npair_square_final}
it is enhanced by  factor of 2 due to the trace over spin. Naively it is also proportional to the number  of flavours $N_f$. This is however not the case since we are interested in production of two quarks of a given flavour. In the real world, the ratio $2/N_c$ is  not a particularly small number, and thus one should not neglect this term off hand.
 However, as it is clear from the 
definition of $\Phi_2$, this term does not depend on the rapidity separation $\eta_1-\eta_2$ 
and thus manifests itself as a pedestal in the  three-particle correlation. 
In what follows we focus on the second term ($\mathtt{B}$).   

In the large $N_c$ limit, the second contribution, proportional to $\Phi_4$, 
dictates that there are only 8 leading $N_c$ contractions for the correlated production.
To understand this consider the trace ${\rm tr} (\tau^a \tau^b \tau^c \tau^d)$.
The color indices are contracted pairwise.  Due to Gaussian averaging of $\rho$, there are two distinct contractions: 
between the nearest neighbours, e.g. 
\begin{equation}
	{\rm tr} (\tau^a \tau^a \tau^c \tau^d)  =  \frac{N_c^2-1}{4 N_c} \delta^{cd}, 
\end{equation}
and the contraction between two matrices separated by the other 
\begin{equation}
	{\rm tr} (\tau^a \tau^b \tau^a \tau^d)  = - \frac{1}{4 N_c} \delta^{bd}.  
\end{equation}
Obviously the latter is suppressed by $1/N_c$; and thus the corresponding contractions of 
the color densities will be ignored.  
This leaves us with 8 possible contractions: there are 4 possible ways to contract a $\rho^f$ 
with one of $\rho^a$,  $\rho^b$,  $\rho^c$,  $\rho^d$; and there are 2 possible ways to pick a 
neighbour to get the leading $N_c$ contribution.  

Therefore in the leading $N_c$, we get 
\begin{align}
	&\langle 
	\rho^a(k) \rho^b(l) \rho^c(\bar{k}) \rho^d(\bar{l})
	\rho^f(m) \rho^f(-m)
    \rangle_P
	\notag 
	\\
	&
	\approx  
	\dimer{a}{k}{f}{m}
	\left( 
	\dimer{b}{l}{f}{-m} \dimer{c}{\bar{k}}{d}{\bar{l}}
	+
	\dimer{d}{\bar{l}}{f}{-m} \dimer{b}{l}{c}{\bar{k}}
	\right)\notag \\
    &
	+ 
	\dimer{b}{l}{f}{m}
	\dimer{c}{\bar{k}}{f}{-m}
	\dimer{d}{\bar{l}}{a}{k}
	+
	\dimer{c}{\bar{k}}{f}{m}
	\dimer{d}{\bar{l}}{f}{-m}
	\dimer{b}{l}{a}{k}
	+ (m\to-m). 
	\label{Eq:Contr}
\end{align}

The final result is symmetric with respect to 
the reversal of the transverse gluon vector $m$. 
To simplify the equations we will keep only the terms we explicitly 
show, the complete expression can be constructed by symmetrizing with 
respect to $m$.  Using a Gaussian model for the projectile
\beq
\left\langle\rho^a(k)\rho^b(p)\right\rangle_P= (2\pi)^2 \mu^2(k)\; \delta^{ab}\;\delta^{(2)}(k+p).
\eeq
we obtain 
\begin{align}
	&
	\frac{1}{(2\pi)^6}
	\langle 
	\rho^a(k)
	\rho^b(l) 
	\rho^c(\bar{k}) \rho^d(\bar{l})
	\rho^f(m) \rho^f(-m)
    \rangle_P
	\notag 
	\\
	&
	\approx 
	\delta^{ab} \delta^{cd} \mu^2(k) \delta^{(2)}(k+m) \mu^2(l)  \delta^{(2)}(l-m)
	\mu^2(\bar{l}) \delta^{(2)}(\bar{k}+\bar{l}) 
	+
	\delta^{ad} \delta^{bc} \mu^2(k)  \delta^{(2)}(k+m) \mu^2(\bar{l})  \delta^{(2)}(\bar{l}-m) 
	\mu^2(l) \delta^{(2)}(l+\bar{k}) 
	\notag \\
    &
	+
	\delta^{cb} \delta^{ad} 
	\mu^2(l) \delta^{(2)}(l+m)
	\mu^2(\bar{k}) \delta^{(2)}(\bar{k}-m)
	\mu^2(k) \delta^{(2)}(\bar{l}+k)
	+
	\delta^{cd} \delta^{ab} 
	\mu^2(\bar{l}) \delta^{(2)}(\bar{l}-m)
	\mu^2(\bar{k}) \delta^{(2)}(\bar{k}+m)
	\mu^2(k) \delta^{(2)}(l+k)\notag \\
	&+ (m\to-m). 
	\label{Eq:ContrMV}
\end{align}
Multiplying by the trace and summing with respect to the color indices we arrive at
\begin{align}
	&\frac{ {\rm tr} (\tau^a \tau^b \tau^c \tau^d)  }{(2\pi)^6}
	\langle 
	\rho^a(k)  \rho^b(l) \rho^c(\bar{k}) \rho^d(\bar{l})
	\rho^f(m) \rho^f(-m)
    \rangle_P
	=  \frac{(N_c^2-1)^2}{4 N_c} \notag \\
	&\times 
	\Big( 
	\mu^2(k)\mu^2(l)\mu^2(\bar{l})   
	\delta^{(2)}(k+m) \delta^{(2)}(l-m)\delta^{(2)}(\bar{k}+\bar{l}) 
	+
	\mu^2(k)\mu^2(\bar{l})\mu^2(l)
	\delta^{(2)}(k+m) \delta^{(2)}(\bar{l}-m)\delta^{(2)}(l+\bar{k}) 
	\notag \\
    &
	+
	\mu^2(\bar{k}) \mu^2(l) \mu^2(k) 
	\delta^{(2)}(l+m)\delta^{(2)}(\bar{k}-m)\delta^{(2)}(\bar{l}+k)
	+
	\mu^2(\bar{l})\mu^2(\bar{k}) \mu^2(k)
	\delta^{(2)}(\bar{l}-m)\delta^{(2)}(\bar{k}+m)\delta^{(2)}(l+k)\notag \\
	&+ (m\to-m). 
	\Big) 
\end{align}

Therefore the correlated piece defined by  the second term of Eq.~\eqref{Npair_square_final}
\bea
\label{Npair_square_corr}
&&\left[ \frac{dN}{d\eta_1d^2pd\eta_2d^2q d\eta_g d^2m } \right]^{\mathtt{B}} _{\rm corr} 
\\ &&=\notag 
- \frac{1}{(2\pi)^4}\frac{g^{10}}{m^2}\int 
d^2k \, d^2\bar{k} \, d^2l \, d^2\bar{l} \;
\langle 
\rho^a(k) \rho^c(\bar{k}) \rho^b(l) \rho^d(\bar{l})
\rho^f(m) \rho^f(-m)
\rangle_P 
{\rm tr}(\tau^a\tau^b\tau^c\tau^d) \Phi_4(k,l,\bar{k},\bar{l};p,q),
\eea
simplifies into 
\begin{align}
	 \notag \left[ \frac{dN}{d\eta_1d^2pd\eta_2d^2q d\eta_g d^2m } \right]^{\mathtt{B}}  _{\rm corr} 
	&= -   \frac{(2\pi)^2 g^{10} \mu^2(m) \mu^2(-m)}{m^2}
	\frac{(N_c^2-1)^2}{4 N_c}
	\\ &\times 
	\int d^2 l \mu^2(l) 
	\Big[ 
\Phi_4(-m,m,-l,l;p,q)
+
\Phi_4(-m,l,-l,m;p,q)
\notag \\ & + 
\Phi_4(l,-m,m,-l;p,q)
+ 
\Phi_4(l,-l,-m,m;p,q)
+ (m\to-m) 
	\Big]
	\label{Eq:CorrN}
\end{align}
which eventually results in 
\begin{align}
	\notag \left[ \frac{dN}{d\eta_1d^2pd\eta_2d^2q d\eta_g d^2m } \right] ^{\mathtt{B}}  _{\rm corr} 
	&= -   \frac{(2\pi)^6 g^{10} \mu^6(m) }{2 m^2}
	\frac{(N_c^2-1)^2}{4 N_c} 
	 \int_0^1 \,
{d\alpha\,d\beta\over (\beta+\bar\beta e^{\eta_1-\eta_2})  (\alpha+\bar\alpha e^{\eta_2-\eta_1})}
\,\notag \\
&\times 
{\rm Tr} \Big[ 
	\delta^{(2)}(p-q)
	\phimerM{} {-m} {p} {\alpha}
	\phimerM{\dagger} {-m} {p} {\alpha}
	\phimerM{} {m} {p} {\beta}
	\phimerM{\dagger} {m} {p} {\beta}
	\notag \\ & 
	+	
   \delta^{(2)}(p-q+2m)
	\phimerM{} {-m} {p} {\alpha}
	\phimerM{\dagger} {m} {q} {\alpha}
	\phimerM{} {m} {q} {\beta}
	\phimerM{\dagger} {-m} {p} {\beta}
	\notag \\ & 
	+	
   \delta^{(2)}(p-q-2m)
	\phimerM{} {m} {p} {\alpha}
	\phimerM{\dagger} {-m} {q} {\alpha}
	\phimerM{} {-m} {q} {\beta}
	\phimerM{\dagger} {m} {p} {\beta}
	\notag \\ & 
	+
	\delta^{(2)}(p-q)
	\phimerM{} {m} {p} {\alpha}
	\phimerM{\dagger} {m} {p} {\alpha}
	\phimerM{} {-m} {p} {\beta}
	\phimerM{\dagger} {-m} {p} {\beta}
\Big]
	\label{Eq:Corr_delta_M}
\end{align}
As expected, this exhibits  a weakening of the correlation 
when the momenta of the quarks are the same (mind the minus sign in front of the integral). Another prominent feature of this expression is that it  is invariant under the reversal of the gluon momentum $m$.

The combination $\phimerM{} {-m} {p} {\alpha} \phimerM{\dagger} {-m} {p} {\alpha}$ 
is proportional to the unit matrix, 
\begin{equation}
	\phimerM{} {m} {p} {\alpha}
	\phimerM{\dagger} {m} {p} {\alpha}
	= \frac{1}{m^4 (\bar{\alpha} p^2 + \alpha (m-p)^2)^2} \left\{ 
		(\bar{\alpha}
		m \cdot p + \alpha m\cdot(m-p))^2
		+ 4 (m\times p)^2
	\right\},
	\label{Eq:Unit}
\end{equation}
while the other relevant combination is given by 
\begin{align}
 \delta^{(2)}(p-q+2m)
	\phimerM{} {-m} {p} {\alpha}
	\phimerM{\dagger} {m} {q} {\alpha}
	&= \frac{ \delta^{(2)}(p-q+2m)}
	{m^4 (\bar{\alpha} p^2 + \alpha (m+p)^2) (\bar{\alpha} (p+2m)^2 + \alpha (m+p)^2)   } 
   \notag \\
	&
	\Big(
	- ( \bar\alpha m\cdot p - \alpha m\cdot(m+p))
	  ( \bar\alpha m\cdot (p+2m) - \alpha m\cdot(m+p)) - 4 (m\times p)^2 
	\notag \\&
	  + 4  i \sigma_3 
	\bar\alpha m^2
	\ m\times p 
 	    \Big)\,. 
	\label{Eq:S3}
\end{align}

Thus
\begin{align}
	&\notag \left[ \frac{dN}{d\eta_1d^2pd\eta_2d^2q d\eta_g d^2m } \right] ^{\mathtt{B}}  _{\rm corr} 
	= -   \frac{(2\pi)^6 g^{10} \mu^6(m) }{2 m^2}
	\frac{(N_c^2-1)^2}{4 N_c} 
\,\notag \\
&\times 
{\rm Tr} \Big[
\delta^{(2)}(p-q)
I_1(\eta_2-\eta_1, -m, p) I_1(\eta_1-\eta_2, m, p)
+
	\delta^{(2)}(p-q+2m)
I_2(\eta_2-\eta_1, m, p) I_2^\dagger(\eta_1-\eta_2, m,  p)
\notag \\ 
&+
	\delta^{(2)}(p-q-2m)
I_2(\eta_2-\eta_1, -m, p) I_2^\dagger(\eta_1-\eta_2, -m, p)
+
\delta^{(2)}(p-q)
I_1(\eta_2-\eta_1, m, p) I_1(\eta_1-\eta_2, -m, p)
\Big]
	\label{Eq:Corr_delta_I}
\end{align}
with
\begin{align}
	I_1(\Delta \eta, m, p) & = 
	\int_0^1 \frac{d\alpha}{\alpha + \bar{\alpha} e^{\Delta \eta}} 
	\phi(m,p,\alpha) \phi^\dagger(m,p,\alpha) , \\
	I_2(\Delta \eta, m, p) & = 
	\int_0^1 \frac{d\alpha}{\alpha + \bar{\alpha} e^{\Delta \eta}} 
	\phi(- m,p,\alpha) \phi^\dagger(m, p +2m,\alpha),  
	\label{Eq:I1_I2}
\end{align}
The integrals $I_1$ and $I_2$ are matrix valued in the spin indices which are traced over in Eq.~(\ref{Eq:Corr_delta_I}).
Explicitly, the definitions of the integrals $I_1$ and $I_2$  read 
\begin{align}
	I_1(\Delta \eta, m, p) & = 
	\int_0^1 \frac{d\alpha}{ \alpha  + e^{\Delta\eta}\bar{\alpha}}  \frac{1}{m^4 (\bar{\alpha} p^2 + \alpha (m-p)^2)^2} \left\{ 
		(\bar{\alpha}
		m \cdot p + \alpha m\cdot(m-p))^2
		+ 4 (m\times p)^2
	\right\},
	\\
	I_2(\Delta \eta, m, p) & = 
    \int_0^1  \frac{d\alpha}{ \alpha  + e^{\Delta\eta}\bar{\alpha}} 
	\frac1
	{m^4 (\bar{\alpha} p^2 + \alpha (m+p)^2) (\bar{\alpha} (p+2m)^2 + \alpha (m+p)^2)   } 
   \notag \\
	&
	\Big(
	- ( \bar\alpha m\cdot p - \alpha m\cdot(m+p))
	  ( \bar\alpha m\cdot (p+2m) - \alpha m\cdot(m+p)) - 4 (m\times p)^2 
	\notag \\&
	  + 4  i \sigma_3 
	\bar\alpha m^2
	\ m\times p 
 	    \Big)\,;
	\label{Eq:I1_I2_s}
\end{align}
The integrals can be computed analytically and the key ingredients are presented in Appendix B.

\begin{figure}
\includegraphics[width=0.5\linewidth]{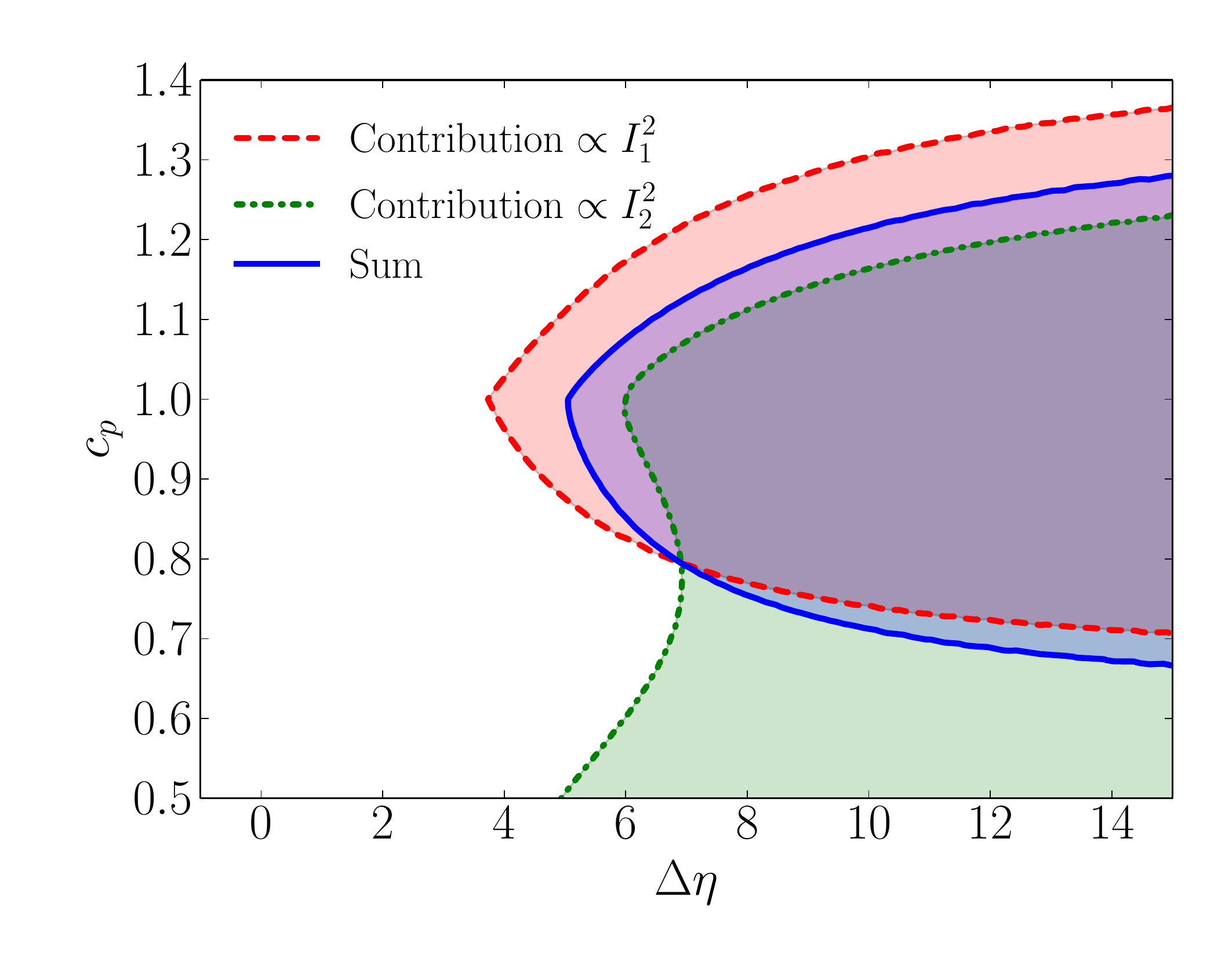}
\caption{ The sign of the correlator as a function of $c_p$ and $\Delta \eta$. 
	The curves illustrate the sign change, or the zeroes of the corresponding contributions. 
	The shaded region show the negative values of the corresponding contributions. 
}
\label{fig:Oz}
\end{figure}

Recall that our goal is to obtain the average of $\cos(\phi_q + \phi_p - 2\phi_m)$, that is 
\begin{equation}
	\gamma  ^{\mathtt{B}}  _{\rm corr} = \langle \cos(\phi_q + \phi_p - 2\phi_m) \rangle ^{\mathtt{B}}  _{\rm corr} 
	= {\cal N}  \int d^2p \int  d^2 q \int d \phi_m  
 \left[ \frac{dN}{d\eta_1d^2pd\eta_2d^2q d\eta_g d^2m } \right] ^{\mathtt{B}}  _{\rm corr}
\cos(\phi_q + \phi_p - 2\phi_m) 
 \label{Eq:gamma}
\end{equation}
Here we have fixed the magnitude of the gluon momentum, while the integral over the quark momenta $p$ and $q$ should performed inside a prescribed momentum bin. Ideally we should choose the size of the two momentum bins to be the same. This can in principle  be done numerically, but would involve performing multi dimensional integrals. To get a qualitative  idea of the behavior of the average we choose a simplified averaging procedure which reduces the problem to a simple two dimensional integral.  
We integrate with respect to 
the absolute value of the momentum of one of the quarks (e.g. $q$) from zero to infinity, while keeping the ratio
of the other quark momentum to the gluon in a finite range. Defining  $\Delta \phi_p = \phi_p-\phi_m$ and $c_p$ = $p/m$ we obtain  
\begin{align}
	\gamma ^{\mathtt{B}}  _{\rm corr} 
& =	 -   {\cal N} 
(2\pi)^7 g^{10} \mu^6(m) 
	\frac{(N_c^2-1)^2}{2 N_c} \notag \\ &\times 
	\int d c_p c_p \int d \Delta \phi_p 
\cos(2\Delta \phi_p)
{\rm Tr} \Big[
I_1(\eta_2-\eta_1, -m, p) I_1(\eta_1-\eta_2, m, p)
\Big]
\notag \\ 
&+
\frac{c_p \cos(2\Delta\phi_p) + 2 \cos(\Delta \phi_p)}{\sqrt{c_p^2+4c_p \cos(\Delta \phi_p) +4}} 
{\rm Tr} \Big[
I_2(\eta_2-\eta_1, m, p) I_2^\dagger(\eta_1-\eta_2, m,  p)
\Big]
\label{Eq:gamma_fin}
\end{align}
 The normalization ${\cal N}$ includes no angular dependence; it is defined through uncorrelated 
production and thus is irrelevant for our qualitative study.

At large $\Delta \eta = \eta_2-\eta_1$, both terms are proportional to $\Delta \eta^2 \exp(- \Delta \eta)$, 
as was shown in Appendix~\ref{Sec:B} 
\begin{align}
	\lim_{\Delta \eta \to \infty} I_1(\Delta \eta, -m, p) I_1(-\Delta \eta, m, p)
	&=  
	\frac{\left[(m \cdot( m  +  p))^2 + 4 (m\times p)^2 \right] \left[ (m \cdot  p)^2 + 4 (m\times p)^2 \right]  }{ m^8 p^4 (m+p)^4} \Delta \eta^2 e^{-\Delta \eta}
	\label{Eq:I1I1lim}
\end{align}
and 
\begin{align}
	\lim_{\Delta \eta \to \infty} I_2(\eta_2-\eta_1, m, p) I_2^\dagger(\eta_1-\eta_2, m,  p)
	&=  
	\frac{
	\left[
(m \cdot( m  +  p))^2 + 4 (m\times p)^2 
	\right] 
	\left[ 
	(m \cdot  p)^2 + 2 m^2\ m\cdot p + 4 (m\times p)^2 
	\right]  
	}{ 
	m^8 p^2 (m+p)^4 (p-2m)^2
   } \Delta \eta^2 e^{-\Delta \eta}. 
 \label{Eq:I2I2lim}
\end{align}
This shows that the rapidity correlations in the projectile wave functions
are quite wide with an exponential decay in rapidity difference  moderated by a power $(\Delta \eta)^2 \exp(-\Delta \eta)$.   
We stress again that this result cannot be immediately confronted to the 
experiment, as the rapidity dependence may be modified by scattering and 
may potentially be further affected by the high energy evolution. 
Let us further analyze both contributions at large rapidity separation,  
in particular we will focus on a more differential observable and fix $c_p$.
Although the angular average of the contribution proportional  
to $I_1^2$ can be performed analytically by the residue analysis, 
the second term in Eq.~\eqref{Eq:gamma_fin} has a branch point singularities and a cut originating from
the square root in the denominator $\sqrt{c_p^2 + 4 c_p \cos \Delta \phi_p  + 4}$ 
and its analytic analysis is complicated. We performed numerical studies of both contributions 
as a function of $c_p$ and $\Delta \eta$ concentrating our attention on the sign change. This is illustrated in Fig.~\ref{fig:Oz}. In the figure,  the shaded regions show where 
the corresponding contributions are negative. We see that the sum has 
negative values concentrated around $c_p \sim 1$.

\begin{figure}
\includegraphics[width=0.5\linewidth]{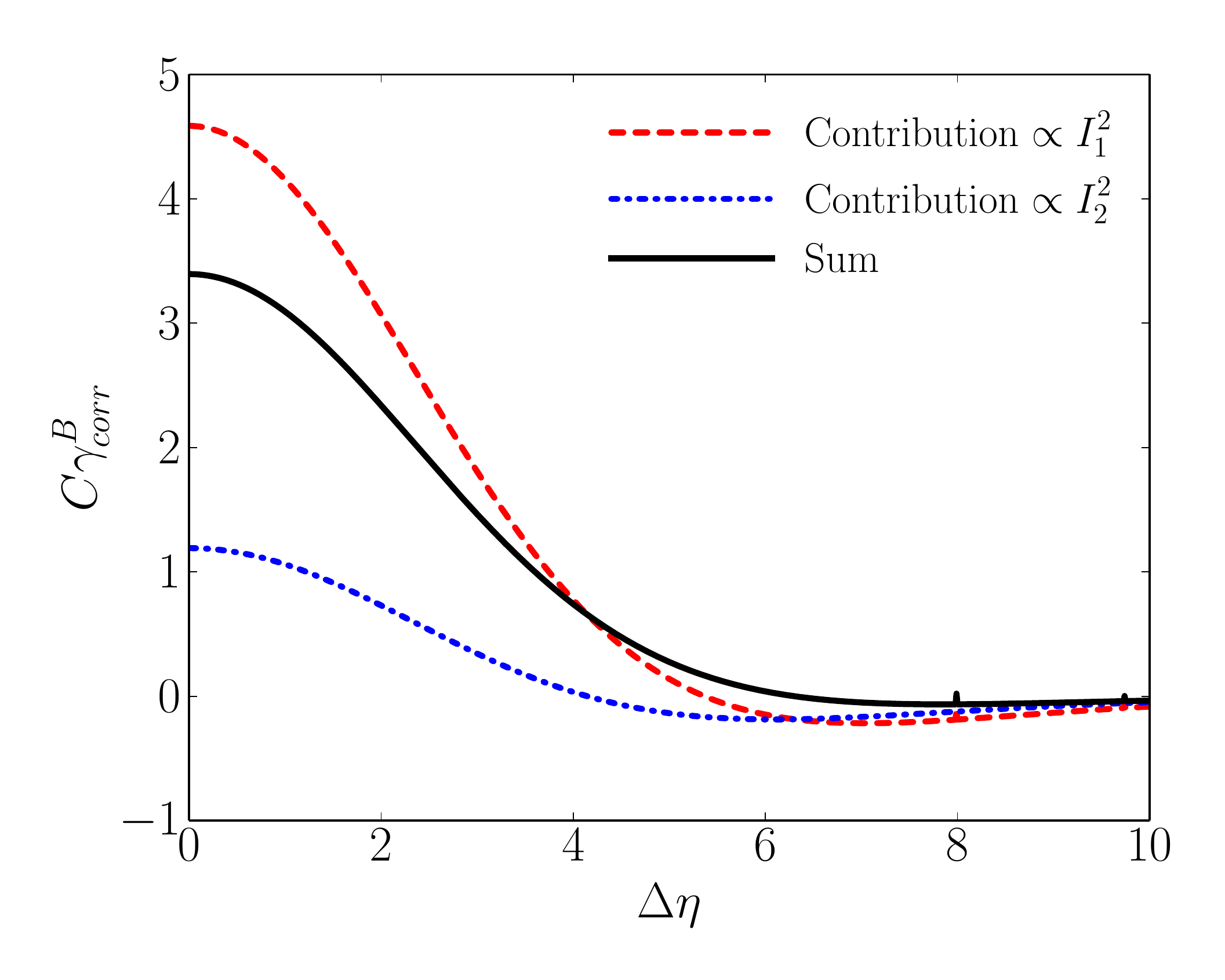}
\caption{   The correlator as a function of $\Delta \eta$. The black solid line is the sum, 
	the red dashed line is the contribution proportional to  from $I_1^2$ 
	and the blue is the part from $I_2^2$. The normalization coefficient $C^{-1} =  
	{\cal N} 
(2\pi)^7 g^{10} \mu^6(m) 
	\frac{(N_c^2-1)^2}{2 N_c} $. 
}
\label{fig:withI2}
\end{figure}

Summing all the terms and integrating numerically in the range $0.9<c_p<1.1$ we get the result illustrated in Fig.~\ref{fig:withI2}. 

\subsection*{Where do the quarks go?}
Although our main goal is to calculate $\gamma^{\mathtt{B}}_{\rm corr}$, it is
instructive to visualize the actual configurations in the wave function that
lead to this result. To this end we plot the different contributions to the
correlation function Eq.~(\ref{Eq:Corr_delta_I}) for different rapidity
differences and different  vales of the ratio $c_p$. 

\begin{figure}
\includegraphics[width=0.5\linewidth]{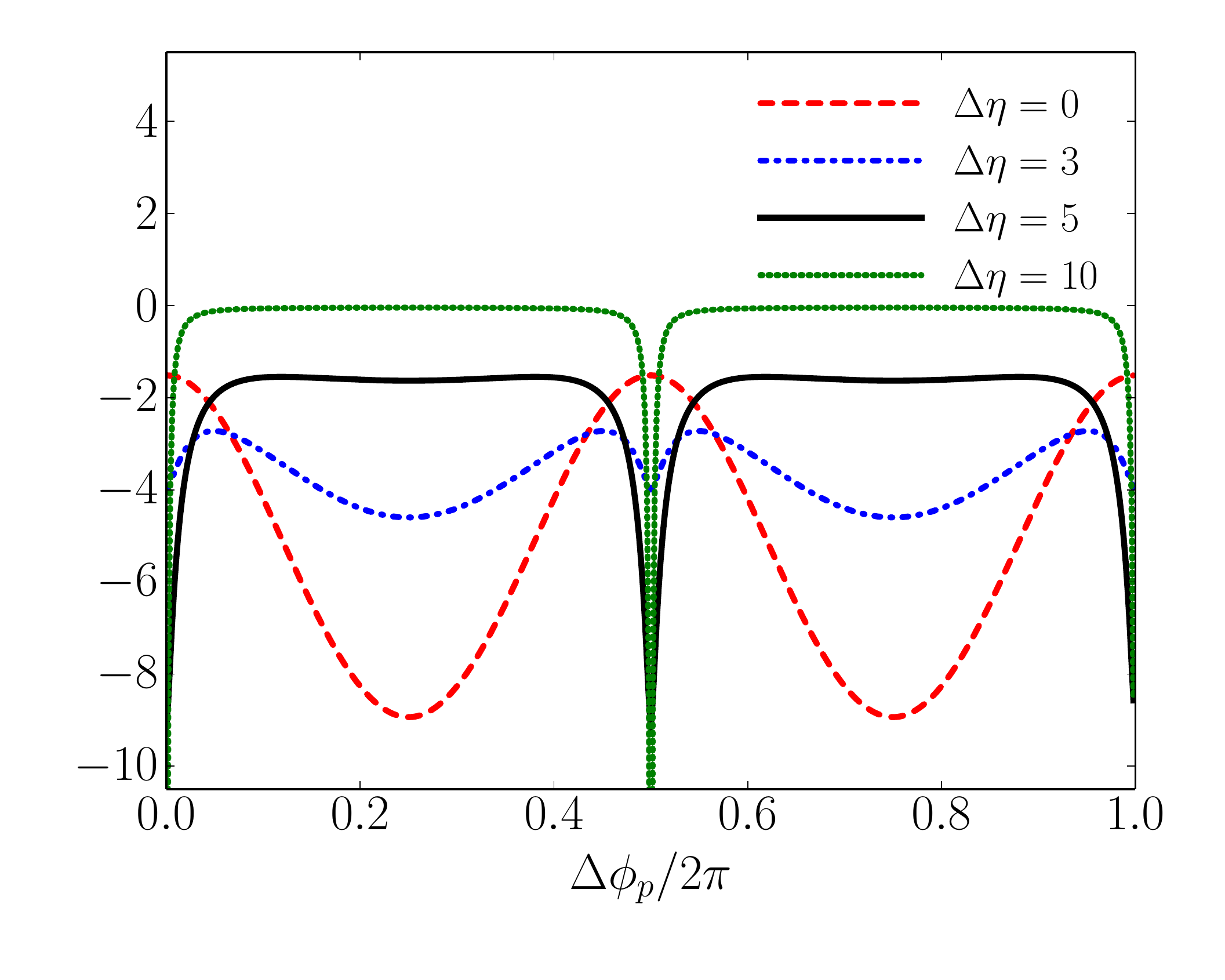}
\caption{ The trace $-\frac{m^4}{2}  {\rm Tr}  I_1(\Delta \eta, -m,p) I_1(- \Delta \eta, m,p) +   m \to -m$ 
as a function of the angle between the momentum of quark $p$ and the momentum of the gluon $m$ for 
$|p|=|m|$ and different $\Delta \eta$. 
}
\label{fig:I1I1}
\end{figure}

Figure~\ref{fig:I1I1} presents the sum of the first and last terms in Eq.~(\ref{Eq:Corr_delta_I}). Recall that in this contribution the transverse
momentum of the second quark $q$ (not shown in the figure) is parallel to $p$.
The figure illustrates that at small rapidity differences $\Delta\eta=0$, the
Pauli blocking most efficiently suppresses configurations where the momenta  of
the two quarks are perpendicular to the momentum of the gluon. The wave
function is thus dominated by the configurations where both quarks are either
parallel or anti parallel with the gluon, naturally leading to positive
contribution to $\gamma^{\mathtt{B}}_{\rm corr}$. At larger rapidity difference the fortunes flip, and
the Pauli blocking becomes stronger for quarks parallel and anti parallel with
the gluon. The wave function becomes dominated by the states where the two
quarks move perpendicular to the gluon in the transverse plane, a configuration
typical of CME. Indeed the sign of $\gamma^{\mathtt{B}}_{\rm corr}$ flips at large rapidity difference.

\begin{figure}
\includegraphics[width=0.29\linewidth]{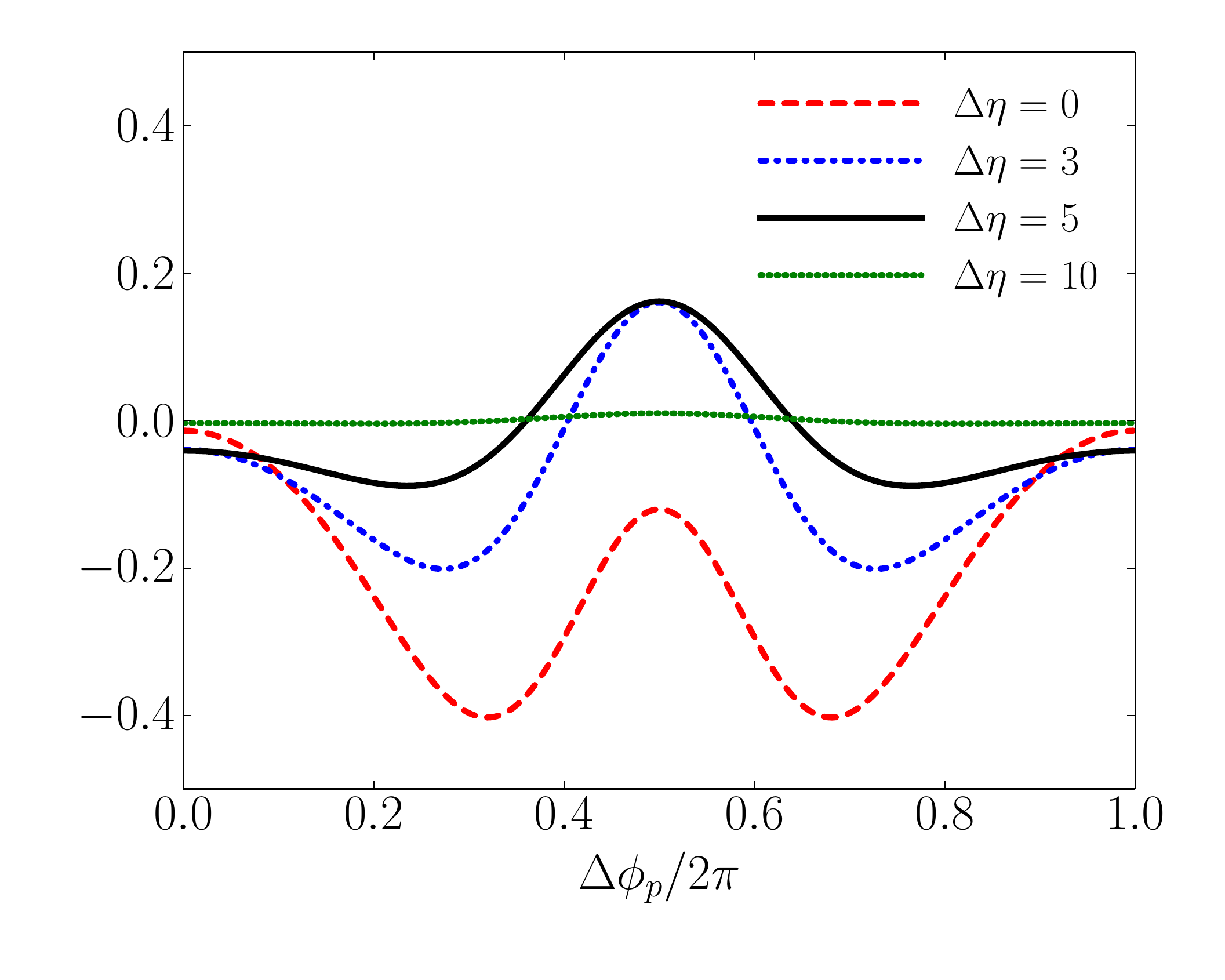}
\includegraphics[width=0.29\linewidth]{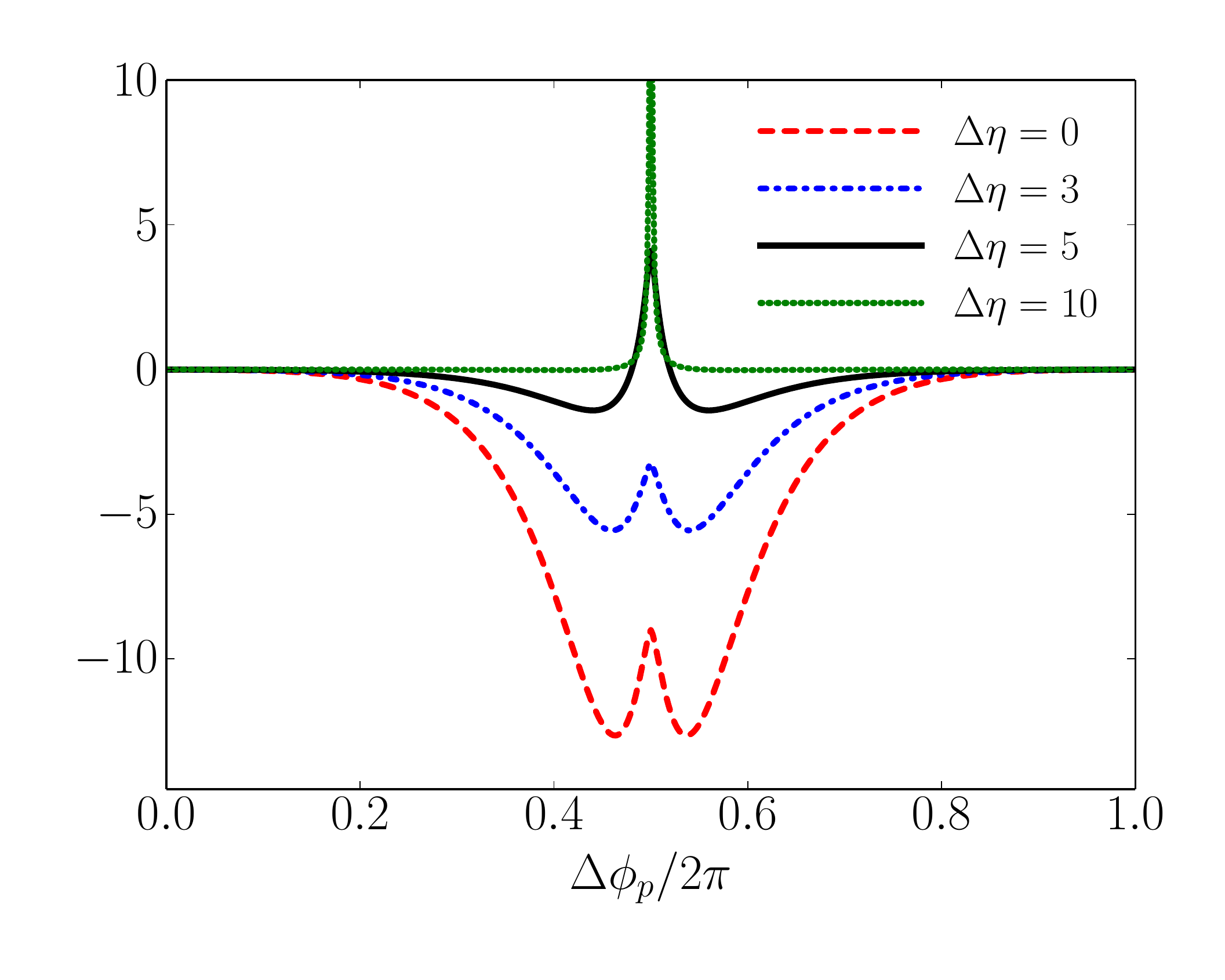}
\includegraphics[width=0.29\linewidth]{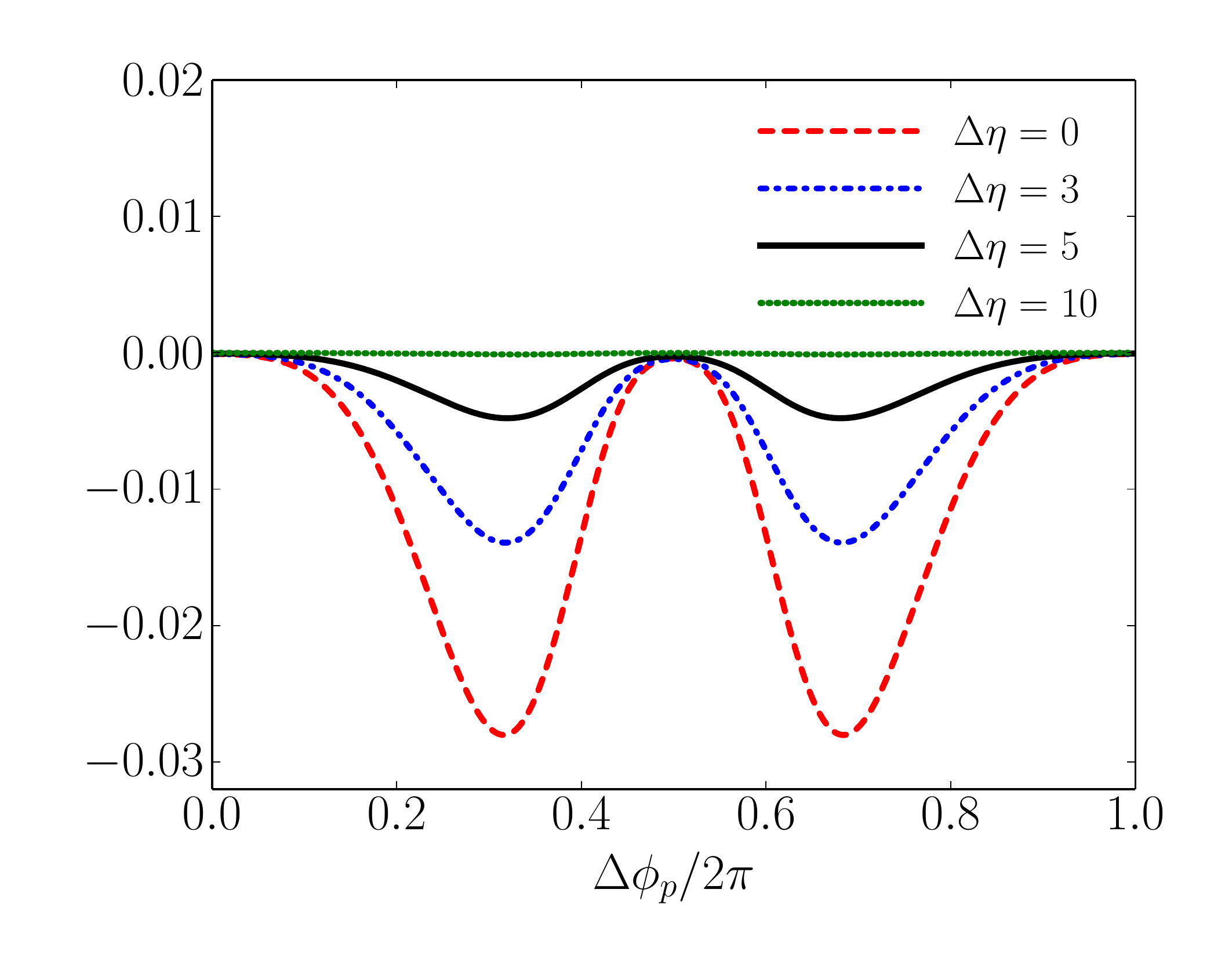}
\caption{ The trace $-\frac{m^4}{2}  {\rm Tr}  I_2(\Delta \eta, m,p) I^\dagger_2(- \Delta \eta, m,p)$ 
as a function of the angle between the momentum of quark $p$ and the momentum of the gluon $m$ for 
$c_p \equiv |p|/|m| = 0.2$ (left), $c_p \equiv |p|/|m| = 1$ (center),  $c_p \equiv |p|/|m| = 5$ (right), and different $\Delta \eta$. 
}
\label{fig:I2I2}
\end{figure}

Figure \ref{fig:I2I2} depicts the second contribution to Eq.~(\ref{Eq:Corr_delta_I}). 
Again we can follow the evolution of the dominant configurations as the function of rapidity difference and also of the ratio of the momenta.

First consider the left panel. Here the magnitude of momentum $p$ is small, and
thus the second quark momentum $q$ is parallel to the gluon momentum $m$ due to
the $\delta$-function in the second term in Eq.~(\ref{Eq:Corr_delta_I}). At
$\Delta\eta=0$, the quark $p$ is mostly parallel or anti parallel with the
gluon, however there is also a significant component of the wave function where
the quark is perpendicular to the gluon. One also observes a symmetry
$\Delta\phi_p\rightarrow \pi-\Delta\phi_p$. Due to this symmetry the
contribution to $\gamma^{\mathtt{B}}_{\rm corr}$ vanishes. As the rapidity difference grows,
the maximum at $\Delta\phi_p=\pi/2$ becomes dominant, but still
 $\gamma^{\mathtt{B}}_{\rm corr}=0$ due to the above mentioned symmetry. In this regime the
dominant configuration is that of a higher momentum quark parallel to the
gluon, and the lower momentum quark perpendicular to the gluon direction.
Finally at very large rapidity difference the distribution becomes flat in the
angle.

The centre panel refers to the value $c_p=1$. At $\Delta\eta=0$ the momentum
$p$ is predominantly either parallel or anti parallel to $m$. The momentum $q$
remains parallel to $m$. The contribution to $\gamma^{\mathtt{B}}_{\rm corr}$ again is very
small due to symmetry. At large $\Delta\eta$ there is also a sharp maximum in
the distribution around $\Delta\phi_p=\pi/2$. In this regime $\gamma^{\mathtt{B}}_{\rm corr}$
is nonvanishing and negative, since $q$ is not strictly parallel to $m$
anymore, but points at acute angle to it.

Finally the right panel is generic for the case  $c_p\gg 1$. Here $q$ and $p$
are approximately  parallel. Here at all $\Delta\eta$ there exist three
preferred configurations: the two quarks are either parallel, anti parallel or
perpendicular to gluon. The sign of $\gamma^{\mathtt{B}}_{\rm corr}$ here depends on the
relative magnitude of the three maxima.

We do not illustrate the  third term in Eq.~(\ref{Eq:Corr_delta_I}), as it is equivalent to the second with $m\rightarrow -m$.

To summarize, we observe that in the regime where we find negative
$\gamma^{\mathtt{B}}_{\rm corr}$ he dominant configurations in he wave function
are very similar to the ones expected due to CME: the momenta of the same
charge quarks are parallel to each other and perpendicular to the momentum of
the gluon, which in our calculation is a proxy to the direction of the event
plane.

\subsection*{The pedestal}

We now turn to the other term in the correlation function, i.e. the first term in Eq.~(\ref{Npair_square_final}). 

\bea
\notag && \left[ \frac{dN}{d\eta_1d^2pd\eta_2d^2q d\eta_g d^2m } \right] ^{\mathtt{A}}  \\ &&= \frac{1}{(2\pi)^4}\frac{g^{10}}{m^2}\int 
d^2k \, d^2\bar{k} \, d^2l \, d^2\bar{l} \;
\langle \rho^a(k) \rho^c(\bar{k}) \rho^b(l) \rho^d(\bar{l})
\rho^f(m) \rho^f(-m)
\rangle_P 
{\rm tr}(\tau^a\tau^b) {\rm tr}(\tau^c\tau^d) \Phi_2(k,l; p) \Phi_2(\bar{k},\bar{l}; q).
\eea

The correlated part in this term comes from eight contractions:
\begin{align}
		&\langle 
	\rho^a(k) \rho^b(l) \rho^c(\bar{k}) \rho^d(\bar{l})
	\rho^f(m) \rho^f(-m)
    \rangle_P
	\notag 
	\\
	&
    = \dimer{a}{k}{f}{m} \dimer{c}{\bar{k}}{f}{-m} \dimer{b}{l}{d}{\bar{l}}
	+
	\dimer{b}{l}{f}{m} \dimer{c}{\bar{k}}{f}{-m} \dimer{a}{k}{d}{\bar{l}}
	\notag \\ &
	+
	\dimer{a}{k}{f}{m} \dimer{d}{\bar{l}}{f}{-m} \dimer{b}{l}{c}{\bar{k}}
	+
	\dimer{b}{l}{f}{m} \dimer{d}{\bar{l}}{f}{-m} \dimer{a}{k}{c}{\bar{k}}
	\notag \\ & +(m\to-m)\,.
\end{align}
Therefore the color summation for this correlators leads to  
\begin{align}
	&{\rm tr} (\tau^a \tau^b)
	{\rm tr} (\tau^c \tau^d)
	\langle 
	\rho^a(k) \rho^b(l) \rho^c(\bar{k}) \rho^d(\bar{l})
	\rho^f(m) \rho^f(-m)
    \rangle_P
	\notag 
	\\
	&
	= 
	\frac{N_c^2-1}{4} (2\pi)^6  \Big(
	\mu^4(m) \mu^2(l)
	\delta ^{(2)}  (k+m) \delta ^{(2)}  (\bar{k}-m) \delta ^{(2)}  (l+\bar{l})
	+
	\mu^4(m) \mu^2(\bar{l})
	\delta ^{(2)}  (l+m) \delta ^{(2)}  (\bar{k}-m) \delta ^{(2)}   (k+\bar{l})
	\notag \\
	&
	+
	\mu^4(m) \mu^2(l)
	\delta ^{(2)}   (k+m) \delta ^{(2)}  (\bar{l}-m) \delta  ^{(2)}  (l+\bar{k})
	+
	\mu^4(m) \mu^2(k)
	\delta ^{(2)}  (l+m) \delta ^{(2)}   (\bar{l}-m) \delta ^{(2)}   (l+\bar{k})
	\Big)+ (m\to-m)\,.
\end{align}
Realizing the $\delta$-functions in the color charge correlators we find
\begin{align}
	 \notag \left[ \frac{dN}{d\eta_1d^2pd\eta_2d^2q d\eta_g d^2m } \right]^{\mathtt{A}}_{\rm corr} 
	&=  \frac{(2\pi)^2 g^{10} \mu^4(m)}{m^2}
	\frac{(N_c^2-1)}{4}
	\\ &\times 
	\int d^2 l \mu^2(l) 
	\Bigg\{\Big[ 
	\Phi_2(-m,l;p)\Phi_2(m,-l;q)
	+
	\Phi_2(-l,-m;p)\Phi_2(m,l;q)
	\notag \\
	&
	+
	\Phi_2(-m,l;p)\Phi_2(-l,m;q)
	+
	\Phi_2(l,-m;p)\Phi_2(-l,m;q)
\Big] \Bigg\} + (m\to -m)\,.
	\label{Eq:CorrN1}
\end{align}
Taking into account that 
\begin{equation}
	\Phi_2(k,l;p) = (2\pi)^2 \delta ^{(2)}  (k-l) \int_0^1 d\alpha\  
	\phi(k,p;\alpha)
	\phi^\dagger(k,p;\alpha) =  (2\pi)^2 \delta^{(2)}(k-l) I_1 (0, k,p) 
	\label{Eq:Phi2M}
\end{equation}
and  realizing the momentum delta functions we arrive to  
\begin{align}
	 \notag \left[ \frac{dN}{d\eta_1d^2pd\eta_2d^2q d\eta_g d^2m } \right] ^{\mathtt{A}} _{\rm corr} 
	&=   \frac{(2\pi)^4 g^{10} \mu^6(m) }{m^2}
	(N_c^2-1) S_\perp
	\\ &\times 
	\Bigg\{ 
		{\rm Tr}\left[ \underbracket{\int_0^1 d\alpha \phi(-m,p;\alpha)\phi^\dagger(-m,p;\alpha)}_{I_1(0,-m,p) }\right] {\rm Tr}\left[ 
		\underbracket{\int_0^1 d\beta\phi(m,q;\beta)\phi^\dagger(m,q;\beta)}_{I_1(0,m,q)} \right]
+(m\rightarrow -m)\Bigg\}\, 
	\label{Eq:CorrN2}
\end{align}
where we introduced the transverse area of the projectile, $S_\perp = (2\pi)^2 \delta^{(2)} (k-k)$.   
As before, we are interested in the following observable 
\begin{equation}
	\langle  \cos(\phi_p + \phi_q - 2 \phi_m) \rangle 
	= 	\langle  \cos(\Delta \phi_p + \Delta \phi_q) \rangle  = 
	\langle  \cos(\Delta \phi_p) \cos(\Delta \phi_q) \rangle -
	\langle  \sin(\Delta \phi_p) \sin(\Delta \phi_q) \rangle 
\end{equation}
To compute the angular average of Eq.~\eqref{Eq:CorrN2}, it is sufficient to compute 
\begin{equation}
	\int d \Delta\phi_q \ I_1(0,m,q)  \cos(\Delta \phi_q) 
\end{equation}
because of the symmetry $	\int d \Delta\phi_q \ I_1(0,m,q)  \sin(\Delta \phi_q) =0$ and the factorization of the angular integrals with respect to 
$\Delta \phi_p$ and $\Delta \phi_q$.  
Before we proceed, we notice that the integral 
\begin{equation}
	\int d \Delta\phi_q \ I_1(0,-m,q)  \cos(\Delta \phi_q)  = 	\int d \Delta\phi_q \ I_1(0,m,q)  \cos(\Delta \phi_q + \pi)  =
	- 	\int d \Delta\phi_q \ I_1(0,m,q)  \cos(\Delta \phi_q) 
\end{equation}
Therefore  we get 
\begin{align}
	\gamma^{\mathtt{A}}_{\rm corr}  &= 
	{\cal N}
	\int d^2q d^2 p \int d\phi_m 
	\left[ \frac{dN}{d\eta_1d^2pd\eta_2d^2q d\eta_g d^2m } \right]^{\mathtt{A}}_{\rm corr} 
 \cos(\phi_p + \phi_q - 2 \phi_m) \notag 
 \\ &=
	- 2
		{\cal N}
	(2\pi)^5 g^{10} \mu^6(m)
	(N_c^2-1) S_\perp m^2 
	\left( \int d {c_p} c_p \int d \Delta \phi_p 
 \, {\rm Tr}  [ I_1(0,m, p) ]\,   \cos(\Delta \phi_p) \right)^2. 
	\label{Eq:gamma1}
\end{align}
Note that naively one might think that the last expression should be multiplied by the number of flavors $N_f$; 
this is however incorrect, as we are interested in the production of same-charge quarks. 
Equation~\eqref{Eq:gamma1} demonstrates  that there is a negative and rapidity-independent contribution to the observable $\gamma$. 

\begin{figure}
\includegraphics[width=0.5\linewidth]{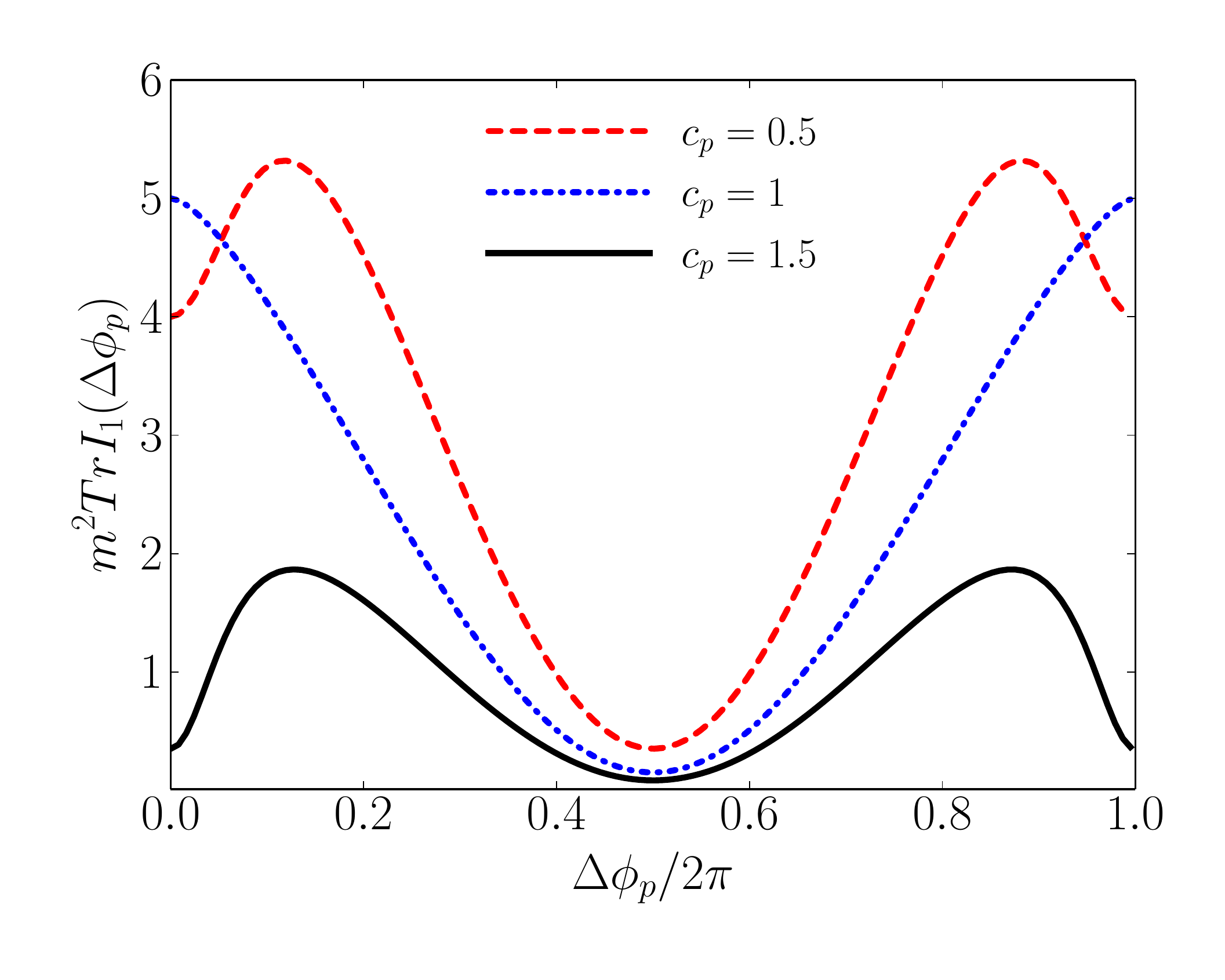}
\caption{ 
	$m^2 {\rm Tr} \ I_1(\Delta \phi_p)$ 
	as a function of $\Delta \phi_p$ for three values of $|p|/|m|=c_p=0.5, 1, 1.5$. 
}
\label{fig:I1}
\end{figure}

To understand the origin of this correlation, it is instructive to consider the  dependence of
$I_1$ on the angle between the gluon and the quark. This dependence is demonstrated in Fig.~\ref{fig:I1}.
As seen from the figure, the momentum of one of the quarks always tends to align with the momentum of the gluon.
This alignment is most favorable for $c_p=1$ and is only approximate otherwise.  The momentum of the other quark on the other hand is anti aligned with that of the gluon, and therefore with the momentum of the first quark. The correlation generating the negative pedestal therefore arises via anti correlation between the momenta of the two quarks mediated by the gluon.

\section{Discussions and Summary} 
\label{Sec:Dis}

In this note we computed three particle ($qqg$) correlations in the projectile wave-function within the McLerran-Venugopalan model. 
In particular we considered the angular average of $\gamma = \langle \cos (\phi_p + \phi_q -2 \phi_m) \rangle $
where $p$ and $q$ are the momenta of the  quarks, and $m$ is the momentum of the gluon. 
We showed that there are two distinct contributions to this quantity: the pedestal, the rapidity-independent
production, with a negative $\gamma$, and the rapidity-dependent contribution originating from Pauli blocking, which 
is characterized by positive $\gamma$ for small $\Delta \eta = \eta_1 -\eta_2$ and negative $\gamma$ for $\Delta \eta \gg 1$. 
The sign change happens at rather large values of $\Delta \eta$ which is not consistent with the experimentally observed 
value of $\Delta \eta \approx 1.5-2$. Nevertheless, qualitatively the rapidity dependence is similar to the experimentally observed one for the 
same charge $\gamma$. We have also seen that in certain kinematics, where
$\gamma^{\mathtt{B}}_{\rm corr}$ is negative, the dominant $qqg$ configurations
in the hadronic wave function have very similar pattern in terms of the
direction of their momenta as expected from CME, even though the physics
producing this pattern is completely different. At the very least this
underscores the necessity to better understand the background for the CME.

To perform more rigorous quantitative studies, one would have to improve on our
calculations in several ways. Most importantly one needs to compute the
scattering with a reasonable model for the target fields.    It would also be
desirable to include more effects of finite density in the projectile wave
function by taking into account the Bogoliubov operator contribution to the
energy evolution \cite{Kovner:2016jfp}. It is not inconceivable that scattering
effects can limit the rapidity range of the Pauli blocking contribution.
Similar effect was observed in Ref.~\cite{Altinoluk:2016vax}, where the
correlated contribution in the wave function was found to decrease at large
rapidity differences as $(\eta_1-\eta_2)^4\exp\{|\eta_2-\eta_1|\}$, while in
the particle production  the decrease was faster -
$(\eta_1-\eta_2)^2\exp\{|\eta_2-\eta_1|\}$. An effect of this type could
shorten the rapidity interval where the quantity $\gamma$ is positive bringing
it closer to the experimental observations. 

Nevertheless we can comment on the quantitative behavior of the two contributions, namely on their dependence on the number 
of colors, the gluon momentum and the projectile transverse area.  
In particular, by comparing the contributions, see Eq.~\eqref{Eq:gamma_fin} and Eq.~\eqref{Eq:gamma1}, we conclude that the term originating from Pauli 
blocking is enhanced by an additional power of $N_c$ and suppressed by the projectile transverse area and the gluon momentum squared $S_\perp m^2$.
Parametrically, both contributions are of the same order if $S_\perp m^2 \sim N_c$. The pedestal-like correlations dominate at large gluon momentum. 

\begin{figure}
\includegraphics[width=0.3\linewidth]{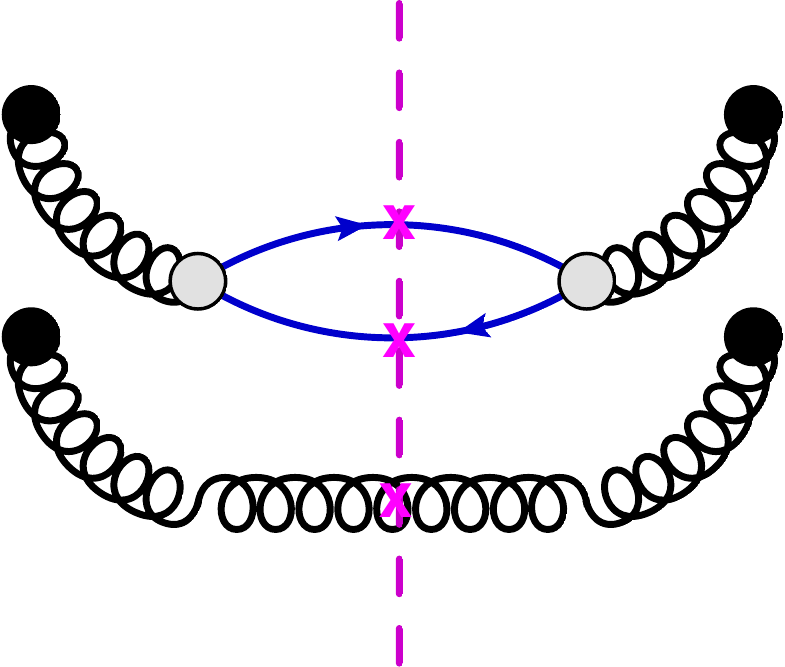}
\caption{The leading order contribution to the three particle correlation involving quark, anti-quark and gluon. The black blobs denote the 
gluon sources, $\rho$. The grey blob in the gluon splitting vertex accounts for the instantaneous interaction too. }
\label{fig:diag}
\end{figure}

Our focus in this paper was on production of same sign (in fact same flavor)
quarks, which upon hadronization are likely to produce same charge pions. 
CME also predicts the behavior of the same charge correlation. In our approach
a reasonable proxy to this quantity should be the $q\bar qg$ correlator. The
behavior of such correlator in the CGC approach is easy to understand
qualitatively.  
  The dominant contribution to such correlator comes from the diagram in Fig.~\ref{fig:diag}, 
  where one of the CGC gluons fluctuates into a quark - anti quark pair.
  The momenta of the quark and anti quark are opposite to each other in the
  frame where the parent gluon has vanishing  transverse momentum. On the other
  hand we know from the calculation of two gluon correlations that the momenta
  of the two gluons are mostly parallel or anti parallel.  We thus expect the
  $q\bar q g$ component of the wave function to be dominated by configurations
  where the quark and anti quark have momenta which are roughly anti parallel
  to each other, and perpendicular to the momentum of the gluon. Such
  configurations are again very similar to the ones expected from CME, and
  would lead to positive $\gamma^{\mathtt{B}}_{\rm corr}$ for opposite charge particles.
  
Our calculations can be extended to study charge-blind correlations  (i.e. without the restriction of the same charge). 
The experimental data in Au-Au collisions at RHIC shows that, the average $\langle \cos(\phi_1+\phi_2-2\phi_3) \rangle$ is negative in a wide range 
of centralities including very peripheral events, see Ref.~\cite{Adamczyk:2017byf}. From the initial state and the CGC perspective, 
an obvious candidate responsible for the explanation of this data would be the three gluon correlation in the 
projectile wave function. However, as it is very well known, see e.g. Ref.~\cite{Kovner:2010xk}, at the leading order  the corresponding number density is 
symmetric under the reversal of any gluon momentum $k\to -k $; this results in vanishing $\langle \cos(\phi_1+\phi_2-2\phi_3) \rangle$ 
at the leading order. Beyond the leading order, as it was demonstrated in Refs.~\cite{McLerran:2016snu,Kovner:2016jfp}, there is no symmetry $k\to -k $. 
At the same order there is also contribution from quarks, which was partially computed in this paper. By combining these pieces together one 
will be able to extract $\langle \cos(\phi_1+\phi_2-2\phi_3) \rangle$ and confront it with the experimentally observed one.     
This is a subject for a separate study.

\begin{acknowledgements}
V.S. thanks L.~McLerran and P.~Tribedy for comments and discussions. 
The research was supported by 
the NSF Nuclear Theory grant 1614640 (A.K.);  Conicyt (MEC) grant PAI 80160015 (A.K.); the  Israeli Science Foundation grants \# 1635/16 and \# 147/12 (M.L.);  the
BSF grants \#2012124 and \#2014707 (A.K., M.L.);  the People Program (Marie Curie Actions) of the European Union's Seventh 
Framework under REA grant agreement \#318921 (M. L) and the COST Action CA15213 THOR (M.L.).
\end{acknowledgements}

\appendix

\section{Light Cone Hamiltonian} \label{sec:A}

In this Appendix we present the Light Cone Hamiltonian calculation of the dressed perturbative state used in Section II. In our notation, see Ref.~\cite{Lublinsky:2016meo}, the light-cone components of four-vectors read $p^\mu\equiv (p^+, p^-,p)$, so $p$ represents the transverse momentum.
The free part of the Light Cone Hamiltonian (LCH, see \cite{Kogut:1969xa,Bjorken:1970ah,Brodsky:1997de})  is
\bea
H_0&=&\int_{k^+>0}{dk^+\over 2\pi}{d^2k\over(2\pi)^2}{k^2\over 2k^+}\ a^{\dagger a}_i(k^+,k)\ a^{a}_i(k^+,k) \\ \nn
\nonumber
&+&\sum_{s}\,\int_{p^+>0}\frac{dp^+ d^2\,p}{(2\pi)^3}\,
\frac{p^2}{2\,p^+} \,\left[d_{\alpha\,s}^\dagger(p^+,p)\,d_{\alpha\,s}(p^+,p)\,+\,\bar d_{\alpha\,s}^\dagger(p^+,p)\,\bar d_{\alpha\,s}(p^+,p)
\right],
\eea
where $a,a^\dagger$ are gluon annihilation and creation operators, $a$ and $\alpha$ are color indices in the adjoint and fundamental representations, respectively, and $i$ and $s$ polarisation and helicity.
This defines the standard free dispersion relations:
\begin{equation}
E_g=k^-={k^2\over 2k^+}, \ \
E_q=p^-={p^2\over 2\,p^+}.
\end{equation}
To zeroth order the vacuum of the LCH is simply the zero energy Fock space vacuum of the operators  $a$, $d$ and $\bar d$:
$$a_q|0\rangle =0,\ \  d_p|0\rangle=0,\ \  \bar d_p|0\rangle=0,\ \
E_0=0.$$
The full Hamiltonian contains several types of perturbations,
\beq\label{deltaH}
\delta H\,=\,\delta H^\rho\,+\,\delta H^{g\,qq}\,+\,\cdots .
\eeq
By $\cdots$ we denote terms that include the soft gluon sector, which is of no relevance for the present work. $\rho$ denotes the color density of the background field, corresponding to the valence or hard degrees of freedom and depending  on transverse coordinates only.

\subsubsection*{Interaction with the background field}
Recall that we are interested in approximate eigenstates of the Hamiltonian in the presence of the background color charge density  due to valence partons. The interaction with the background charge is comprised of three terms
\beq
\delta H^{\rho}\,=\,\delta H^{\rho\,g}\,+\, \delta H^{\rho\,qq}\,+\, \delta H^{\rho\,gg}\ .
\eeq
The last term is of no interest to us since it does not involve quarks. The remaining ones are
\bea
\delta H^{\rho\,g}&=&
\int_0^\infty {dk^+\over 2\pi}{d^2k\over (2\pi)^2}{g\,k_i\over \sqrt{2}\,|k^{+}|^{3/2}}\
\Big[a^{\dagger a}_i(k^+,\,k)\  \rho^a(-k) 
+ a^a_i(k^+,\,k)\,\  \rho^a(k)\Big] , \\
\delta H^{\rho\,qq}&=&\sum_{s}
\int {dk^+d^2k\,dp^+d^2p\over (2\pi)^6}{g^2\over \,(k^{+})^2}\
\Big[d^{\dagger}_{\alpha\,s}(p^+,\,p)\ \tau^a_{\alpha\beta}\ \bar d^{\dagger}_{\beta\,s}(k^+\,-\,p^+,\,k-p)\
\rho^a(-k)
+   h.c.\Big].
\eea

\subsubsection*{Quark-gluon interaction}

The quark-gluon interaction responsible for quark production reads
\bea
\delta H^{g\,qq}&=& g\,\tau^a_{\alpha\beta}\,\sum_{s_1,s_2}\,\int {dp^+\,d^2p\, dk^+\,d^2k
\over 2^{3/2}\,(2\pi)^6\,(k^+)^{1/2}}\,\theta(k^+\,-\,p^+)\,\Gamma^i_{s_1\,s_2}(k^+,k,p^+,p)\nn\\
&\times& \Big[{a}_{i}^a(k^+,k)\,d^\dagger_{\alpha,\,s_1}(p^+,p)\,\bar
d^\dagger_{\beta,s_2}(k^+-p^+,k-p)
+
h.c.\Big], \eea
with the vertex $\Gamma^i$ defined as
\bea
\Gamma^i_{s_1s_2}(k^+,k,p^+,p)&=&\chi_{s_2}^\dagger\left[2\frac{k_i}{k^+}-
\frac{\sigma\cdot p}{p^+}\sigma^i-
\sigma^i\frac{\sigma\cdot (k-p)}{(k^+-p^+)}\right]\chi_{s_1} \\
&=&\chi_{s_2}^\dagger\left[2\frac{k_i}{k^+}-\left({p_i\over p^+}+{k_i-p_i\over k^+-p^+}\right)
+i\epsilon^{im}\sigma^3\left({p_m\over p^+}-{k_m-p_m\over k^+-p^+}\right)\right]\chi_{s_1} \nn \\
&=&\delta_{s_1s_2}\left[2\frac{k_i}{k^+}-\left({p_i\over p^+}+{k_i-p_i\over k^+-p^+}\right)
+2is_1\epsilon^{im}\left({p_m\over p^+}-{k_m-p_m\over k^+-p^+}\right)\right],\nn
\eea
and the spinors $\chi_{s=1/2}=(1,0)$ and $\chi_{s=-1/2}=(0,1)$. 

Diagonalising the perturbation $\delta H$ in (\ref{deltaH}) perturbatively leads to the wavefunctions (\ref{vd}), (\ref{v4}).
More detailed calculations could be found in ~\cite{Altinoluk:2016vax,Lublinsky:2016meo}.

\section{Integrals}
\label{Sec:B} 
Here we list integrals essentials to compute $I_1$ and $I_2$, defined in Eq.~\eqref{Eq:I1_I2}. 
We start from $I_1$. 
\begin{equation}
	\int_0^1 \frac{d\alpha}{\alpha + \bar{\alpha} e^{\Delta\eta} } \frac{1}{ (\bar{\alpha} p^2 + \alpha (m-p)^2)^2}  = 
	\frac{(m-p)^2-p^2}{e^{\Delta\eta} (m-p)^2-p^2} \frac{1}{p^2(p-m)^2 }
	+
	\frac{e^{\Delta\eta} -1 } {(e^{\Delta\eta} (m-p)^2-p^2)^2} \left[ \ln \frac{(m-p)^2}{p^2} +\Delta\eta \right]
\end{equation}

\begin{align}
	\int_0^1&
\frac{d\alpha}{\alpha + \bar{\alpha} e^{\Delta\eta} } 
 \frac{(\bar{\alpha}
		m \cdot p + \alpha m\cdot(m-p))^2
	}{(\bar{\alpha} p^2 + \alpha (m-p)^2)^2} =  \frac{(p-m)^2-p^2}{e^{\Delta\eta}(p-m)^2-p^2} \left(\hat p \cdot \widehat{(p - m) }  \right)^2
+ 
\frac{1}{(e^{\Delta\eta}(p-m)^2-p^2)^2}
\notag \\&
\times	 \left[ 
p\cdot(p-m) \left\{ p \cdot (p+m) 
-
e^{\Delta \eta} (p-m) \cdot(p-2m)
\right\} \ln \frac{(p-m)^2}{p^2}
+ (m\cdot p - e^{\Delta\eta} m\cdot(m-p))^2 \frac{\Delta\eta}{e^{\Delta \eta}-1}
\right]
\end{align}
Using the identity 
\begin{equation}
	 \left(\hat p \cdot \widehat{(p - m) }  \right)^2 = 1 - \frac{(m\times p)^2}{p^2 (p-m)^2} 
\end{equation}
we obtain 
\begin{align}
	& m^4 I_1(\Delta\eta , m, p) =
 4 (m\times p)^2 	\frac{e^{\Delta\eta} -1 } {(e^{\Delta\eta} (m-p)^2-p^2)^2} \left[ \ln \frac{(m-p)^2}{p^2} + \Delta\eta \right]
 + \notag \\&
 \frac{(m-p)^2-p^2}{e^{\Delta\eta} (m-p)^2-p^2}
\left(1 + 3  \frac{(m\times p)^2}{p^2(p-m)^2} 
 \right) 
+ 
\frac{1}{(e^{\Delta\eta}(p-m)^2-p^2)^2}
\notag \\ & 
\times
\left[ 
p\cdot(p-m) \left\{ p \cdot (p+m) 
-
e^{\Delta \eta} (p-m) \cdot(p-2m)
\right\} \ln \frac{(p-m)^2}{p^2}
+ (m\cdot p - e^{\Delta\eta} m\cdot(m-p))^2 \frac{\Delta\eta}{e^{\Delta \eta}-1}
\right]
\end{align}
In the limiting case of $\Delta\eta \to \infty$
\begin{equation}
	\lim_{\Delta \eta \to \infty} I_1(\Delta\eta, m, p) = 
	 \frac{1}{m^4} \frac{(m \cdot( m  -  p))^2 + 4 (m\times p)^2}{(m-p)^4} \Delta \eta e^{-\Delta \eta}
\end{equation}
and 
\begin{equation}
	\lim_{\Delta \eta \to \infty} I_1(- \Delta\eta, - m, p) = 
	 \frac{1}{m^4 p^4} \left( (m \cdot  p)^2 + 4 (m\times p)^2 \right) \Delta \eta. 
\end{equation}

The relevant integrals for $I_2$ are 
\begin{align}
&	\int _0^1 \frac{d\alpha}{\alpha + \bar{\alpha} e^{\Delta\eta} } 
 		\frac{1}
		{
		 (\bar{\alpha} p^2 + \alpha (m+p)^2) (\bar{\alpha} (p+2m)^2 + \alpha (m+p)^2)  
		}\notag \\
		& = 
		\frac{(e^{\Delta\eta} -1 ) \Delta \eta}
		{(e^{\Delta \eta} (p+m)^2-p^2)(e^{\Delta\eta} (p+m)^2 - (p+2m)^2)}
		\notag \\ & +
		\frac{1}{
			( (p+2m)^2 - p^2) (p+m)^2
		}\notag \\ & \times
		\Bigg(
		\frac{(m+p)^2-p^2}{e^{\Delta\eta}(m+p)^2-p^2} \ln \frac{(p+m)^2}{p^2}
		+
		\frac{(m+p)^2-(p+2m)^2}{e^{\Delta\eta}(m+p)^2-(p+2m)^2} \ln \frac{(p+2m)^2}{(p+m)^2}
		\Bigg), 
\end{align}
\begin{align}
&	\int_0^1  \frac{d\alpha}{\alpha + \bar{\alpha} e^{\Delta\eta} } 
 		\frac{ ( \bar\alpha m\cdot p - \alpha m\cdot(m+p))
	  ( \bar\alpha m\cdot (p+2m) - \alpha m\cdot(m+p)) }
		{
		 (\bar{\alpha} p^2 + \alpha (m+p)^2) (\bar{\alpha} (p+2m)^2 + \alpha (m+p)^2)  
		}\notag \\
		&=
		- \frac{\Delta\eta}{e^{\Delta\eta}-1}
		\frac{
			(m\cdot p + e^{\Delta\eta} (m^2+m\cdot p))
			(m^2 + (1+e^{\Delta\eta}) (m^2+m\cdot p))
		}{
		( e^{\Delta\eta} (m+p)^2-p^2 )
		( (p+2m)^2-  e^{\Delta\eta} (m+p)^2 )
		}
		\notag \\ & + 
		\frac{(p^2+m\cdot p)( 2(m^2+m\cdot p)^2 + m^2 m\cdot p + p^2(3m^2+2m\cdot p) )}
		{
			(e^{\Delta\eta} (p+m)^2-p^2)
			( (p+2m)^2 - p^2) (p+m)^2
		} \ln \frac{(p+m)^2}{p^2}
		\notag \\ & -
		\frac{ (p+m)\cdot(p+2m) [  (p+2m)^2 m\cdot (p+m) + m\cdot p (p+m)^2  ] }
		{
			(e^{\Delta\eta} (p+m)^2-(p+2m)^2)
			( (p+2m)^2 - p^2) (p+m)^2
		}
	 \ln \frac{(p+2m)^2}{(p+m)^2}
\end{align}
and, finally,
\begin{align}
&	\int_0^1  \frac{d\alpha}{\alpha + \bar{\alpha} e^{\Delta\eta} } 
 		\frac{\bar{\alpha}}
		{
		 (\bar{\alpha} p^2 + \alpha (m+p)^2) (\bar{\alpha} (p+2m)^2 + \alpha (m+p)^2)  
		}
		= \frac{\Delta \eta}{(e^{\Delta \eta} (m+p)^2 -p^2 )  ((p+2m)^2 -  e^{\Delta \eta} (m+p)^2 )} 
		\notag \\ & + 
		 \frac{1}{ (e^{\Delta \eta} (m+p)^2 -p^2 )  ( (p+2m)^2 - p^2) } \ln \frac{(p+m)^2}{p^2}
		\notag \\ & + 
		 \frac{1}{ ( (p+2m)^2 -  e^{\Delta \eta} (m+p)^2 )  ( (p+2m)^2 - p^2) }  \ln \frac{(p+m)^2}{(p+2m)^2}. 
\end{align}
Multiplying the integrals by the corresponding factors, one may obtain $I_2$. The resulting equation is cumbersome; we 
refrain from providing its explicit result here. 

Nevertheless, in the limit of large $\Delta \eta$ we obtain  
\begin{equation}
	\lim_{\Delta \eta \to \infty} I_2(\Delta\eta, m, p) = 
	 -\frac{1}{m^4} \frac{(m \cdot( m  +  p))^2 + 4 (m\times p)^2}{(m+p)^4} \Delta \eta e^{-\Delta \eta}
\end{equation}
and 
\begin{equation}
	\lim_{\Delta \eta \to \infty} I_2(- \Delta\eta, m, p) = 
	-  \frac{1}{m^4 p^2 (p-2m)^2} \left( (m \cdot  p)^2 + 2 m^2\ m\cdot p + 4 (m\times p)^2 \right) \Delta \eta. 
\end{equation}

\bibliography{cmecgc}

\end{document}